\documentclass[aps,showpacs,preprintnumbers,amsmath,amssymb,superscriptaddress]{revtex4}

\usepackage{graphics}
\usepackage{epsfig}
\usepackage{color}
\newcommand{\beq}{\begin{equation}}
\newcommand{\eeq}{\end{equation}}

\newcommand{\drv}[2]{\frac{\partial  #1}{\partial  #2}   }
\newcommand{\opdrv}[1]{\frac{\partial }{\partial  #1}  }

\newcommand{\ddrv}[2]{\frac{\partial ^{2}  #1}{{\partial  #2}^2}   }

\def\la{\hbox{\raise.35ex\rlap{$<$}\lower.6ex\hbox{$\sim$}\ }}
\def\ga{\hbox{\raise.35ex\rlap{$>$}\lower.6ex\hbox{$\sim$}\ }}

\def\beq{\begin{equation}}
\def\eeq{\end{equation}}
\def\beqa{\begin{eqnarray}}
\def\eeqa{\end{eqnarray}}
\def\bseq{\begin{subequations}}
\def\eseq{\end{subequations}}

\def\order#1{{\cal O}\left({#1}\right)}

\def\plottwo#1#2{\centering \leavevmode
\epsfxsize=0.45\columnwidth \epsfbox{#1} \hfil
\epsfxsize=0.45\columnwidth \epsfbox{#2}}

\definecolor{darkgreen}{rgb}{0,0.5,0}

\begin{document}


\title{Global transient dynamics of three-dimensional hydrodynamical disturbances
in a thin viscous accretion disk  }
\author{P. Rebusco}
\affiliation{Kavli Institute for Astrophysics and Space Research, MIT,
Cambridge, MA}
\email[]{pao@space.mit.edu}
\author{O.M. Umurhan}
\affiliation{Astronomy Unit, School of Mathematical Sciences, QMUL,\,
Mile End Road, London E1 4NS, UK}
 \affiliation{Department of Astronomy, City
College San Francisco, CA 94112, USA}
\email[]{umurhan@maths.qmul.ac.uk}
\author{W. Klu\'{z}niak}
\affiliation{Institute of Astronomy, Zielona G\'ora University, ul. Lubuska 2, 65-265 Zielona G\'ora, Poland}
\affiliation{Nicolaus Copernicus Astronomical Center, ul. Bartycka 18, 00-716 Warsaw, Poland}
\email[]{wlodek@camk.edu.pl}
\author{O. Regev}
\affiliation{Department of Physics, Technion-Israel Institute of
Technology, 32000 Haifa, Israel}
\affiliation{Dept of Astronomy, Columbia University, New York NY, 10027}
\email[]{regev@astro.columbia.edu}

\date{\today}

\begin{abstract}
{Thin viscous Keplerian accretion disks are considered
asymptotically stable, even though they can show significant
 dynamic activity on short timescales.
In this paper the dynamics of non-axisymmetric hydrodynamical disturbances
of  disks are investigated analytically building upon the steady state three-dimensional structure and evolution of axisymmetric perturbations explored in previous work.
Assuming a polytropic equation of state solutions are found by means of an
asymptotic expansion in the small parameter measuring
the ratio of the disk thickness to characteristic radius.
In-depth analysis shows that every perturbation that
disturbs the radial velocity induces significant transient growth
in the (acoustic) energy of the evolving disturbance. This effect is most evident in the
density and vertical velocity.   The transient growth observed is tied to the non-separable
nature of the solutions where, in particular, pattern evolution is controlled by a
similarity variable composed of the radial coordinate and time.
This leads to  growing winding  perturbations that display successive radial peaks and troughs.
We argue that these transient non-axisymmetric structures
may precipitate secondary instabilities which, consequently,
may be a critical element for a new alternative picture of turbulence arousal
in non-magnetized astrophysical disks. }

\end{abstract}

\pacs{Valid PACS appear here}
\maketitle
\section{Introduction}

Since the late 1980s a new perspective has developed in hydrodynamic stability theory.
This new paradigm arose from the long-standing problem of linearly stable
shear flows that experimentally exhibit transition into turbulence.
The conventional approach had been to examine the linear stability
of fluid systems via modal analysis (``normal-modes"). In practice this means that the
determination of the eigenvalues and eigenfunctions of the linearized
perturbation equations of a given flow indicates the time asymptotic behavior of
the disturbances and, consequently, helps to determine the long-time stability
of the base flow.  One of the best expositions of this
classical approach is found in the book by Drazin \& Reid \cite{DR}.
\par
The origin of the new perspective can be traced to the fact that
linear stability analysis of shear flows in general
gives rise to {\em non-normal} operators, i.e.,
linear operators that do not commute with their
adjoint.  A typical modal analysis of a problem governed by a
non-normal operator can lead to an incomplete description of
the full breadth of responses possible for the
original {\em initial-value problem} (IVP).
For example, non-normal operators will
have eigenfunctions that are non-orthogonal, and/or imply the existence of solutions
which are unobtainable analytically (see the discussion in \cite{CRI} Chapter 8).\par
Especially critical in this matter is the fact that the non-orthogonality of the eigenmodes
can lead to {\em transient dynamics} which, in turn, can imply strong {\em transient growth} (TG)
for suitable initial conditions, e.g., in perturbation energy or enstrophy.
The implications of this for a variety of shear flows has been studied in numerous
earlier works, e.g., \cite{BB,G,BFa}.
The framework for stability calculations has thus
shifted from just focusing on the time-asymptotic behavior of a disturbance
towards studying its TG as well.
The recent review article by Schmid \cite{SR}
gives an up-to-date account on non-modal stability theory and its successes.
A detailed exposition of the subject, including
its different aspects and possible extensions, can be found in the book by
Schmid \& Hennigson \cite{SH}.

Ioannaou \& Kakouris \cite{petros}
were first to apply this non-modal perspective for an astrophysical setting, namely,
for the problem of accretion disks (``ADs" and ``AD" for singular usage).
Accretion disks are both important and ubiquitous astrophysical
objects which are thought to power systems as diverse as young stellar objects, close binary
systems and active galactic nuclei.
ADs are flattened, swirling flows in a gravitational field of a central
compact object, wherein high specific angular momentum fluid accretes onto the central object.
In order to reconcile theoretical models with observations,
an efficient mechanism is needed that dissipates energy and transports angular momentum
because the slow spiraling-in and eventual accretion of fluid
depends critically on this process.
When ``viscous" ADs were theoretically proposed\cite{SS,LBP}
it was recognized that an anomalous dissipation and transport mechanism
must be present in ADs since their hydrodynamical
Reynolds numbers (${\rm Re}$) are enormous.
Because fluid turbulence greatly enhances transport,
turbulence has been
proposed to operate in ADs as angular momentum
can be transported in rotating flows with the help of a turbulent
eddy viscosity.   To date, a detailed
theoretical understanding of turbulence and the transition to it
is still lacking.  Consequently, the effective viscosity in disks
has been approached in a phenomenological way through
parameterizing the effective viscosity coefficient with the help of
a non-dimensional parameter($\alpha$) on the basis of
dimensional arguments.
This simple approach has been exceptionally
fruitful, giving rise to successful interpretations of many basic
observational results \cite{LP,FKR}.\\

\par Robust hydrodynamical stability criteria like the Rayleigh
and Solberg-Hoiland criteria indicate that thin non-magnetized Keplerian
ADs are linearly stable.
To date there are no demonstrations of long-time dynamical activity in
global hydrodynamical simulations of AD flows with sufficiently high {\rm Re}.
On the other hand,  high resolution
3D numerical simulations of local disk sections \cite{LL}
report that a {\em subcritical} transition to turbulence does exist at very high
Reynolds numbers despite suggestions that Coriolis effects quench such
a transition\cite{BHS,HBW}.  However,
the efficiency of turbulent transport in such subcritical flows
appears to be insufficient to explain the transport implied by the observations of ADs. In this study
we use the term {\em global} when
the calculation includes a sizable portion of the AD. In contrast, a {\em local calculation}
is one that is based on the so-called {\em shearing box} approximation \cite{GLB,BH,UR}
in which calculations are performed on small ``Cartesionized"
sections of the disk.

In the years following  the work of Ioannaou \& Kakouris \cite{petros} (who had employed
a global approach albeit to a two-dimensional configuration) a number of research groups have used
the local approximation to study the relevant IVP in various settings \cite{yecko,UR,asaf,CZTL,TCZCL,TCZ,AMR,MAN}.  A common conclusion reached by these studies is that
TG may be copious and that, under the right conditions,
nonlinear interactions {\em may} give rise to what is called
a {\em bypass} transition to turbulence. It is worth mentioning that vortices and spiral waves are
singled out in some of these works as being instrumental for such a transition.



This work is intended to further understanding
what the non-modal IVP perspective can teach us about the global
transient dynamics of thin slightly viscous hydrodynamical Keplerian disks.
To be specific, we consider flows with a  viscosity (perhaps of turbulent origin\cite{LL}) that is
{\em insufficient} to provide the angular momentum transport implied by observations.
We are interested in examining the excitation of global secondary flows that can, in turn, possibly
give rise to angular momentum transport and/or
create conditions for a secondary instability atop the weakly turbulent state
\cite{kato78,kleiber,latter,kersale,Shaviv-Umurhan}.

In the context of ADs one is confronted with a vast system
on which there is no experimental control.
In lieu of this, Ionnaou \& Kakouris \cite{petros} reported in their study that
stochastic forcing was found to lead to persistent activity with angular momentum
transported outward. More recently, Zhuravlev \& Shakura \cite{Shakura} applied
optimal perturbation strategy to two-dimensional sub-Keplerian toroidal
configurations.  They found that optimal perturbations giving rise to substantial
TG are composed of certain
combinations of non-axisymmetric eigenmodes.

Our tactic is to study the transient dynamics of {\em specific} three-dimensional
perturbations of a Keplerian AD
in which the vertical structure is taken into account. The base flow is
the Klu\'zniak-Kita \cite{KK} (hereafter KK) analytical solution of a polytropic steady
viscous axisymmetric disk. This solution was obtained
by representing all dependent functions as an asymptotic series
in the small parameter $\epsilon$
which measures the disk's ``thinness"
(i.e., the ratio of the disk's vertical thickness to its typical radial scale).
The lowest order solution is the classical
Shakura-Sunyaev solution \cite{SS}, while higher order terms provide the velocities and
structure functions in the meridional plane, as well as corrections to the Keplerian angular velocity
\cite{footnote1}.
This kind of perturbation strategy for ADs had first been introduced
in the context of AD inner boundary layer\cite{R}.
We wish to reiterate here that implicit in the KK solution
is that there is some viscosity on the smallest scales,
perhaps due to the subcritical transition discussed previously
\cite{LL}.
\par
Umurhan  et al. \cite{UNRS}
(hereafter UNRS), extended the analytical KK solution to consider its
time dependent axisymmetric response.
The analysis of the corresponding IVP indicated that
typical initial data naturally exhibits transient growth
lasting many rotation periods before ultimately decaying.  The lifespan of this response
 is inversely proportional to the viscosity parameter $\alpha$, which is consistent for
 studies of simpler systems \cite{SH}.
 Because UNRS examined
only axisymmetric initial disturbances and flows, the transiently growing patterns could not induce
any effective radial angular momentum transfer. In addition, no obvious mechanism was
identified that could serve as
a suitable secondary instability candidate.

In this paper we report on a work based on the same ideas applied to the same slightly viscous
system, but allowing for {\em non-axisymmetric} disturbances.
Consequently the results, which remain {\em analytical} as in UNRS, exhibit a far richer structure.
The asymptotic expansion procedure used in UNRS is applied in which the
same small parameter (i.e., the disk's 'thinness') is exploited resulting in an IVP that is
analytically treatable.  The expansion procedure employed here develops a finite-amplitude
nonlinear solution.  Thus, although the higher order terms and time-dependencies are governed
by linear operators, the solutions developed are finite-amplitude
and they should not be confused as being infinitesimal solutions.
\par
The use of approximation methods and the simplistic polytropic assumption have  the obvious  advantages  that the
treatment can be analytical and the responsible physical effects leading to any interesting dynamics may be
transparently traced. It is clear, however, that the present analytic analysis ultimately should be complemented
with a detailed and uncompromising numerical solution which includes a proper treatment of energy generation and transfer\cite{RG}.
We also stress that we do not aim to obtain a {\em general} solution for the IVP, nor do we
intend to itemize all
possible solutions. We also do not aim to find the {\em optimal} perturbation, which
would be a strategy appropriate for a laboratory transition study.
The purpose of this study is to show that there {\em exists} a solution
 that gives rise to prominent transient growth in the energy when small {\em but non-infinitesimal}
 disturbances are introduced.
 We also investigate how the different parameters of the system
(i.e., $\alpha$, symmetry) affect this growth.
Implicit in our approach is the tacit assumption that the
possible perturbation spectrum in a realistic AD is so rich as to allow essentially any initial condition we desire.
We are specifically interested in finding a global transiently growing spatial pattern
of the density and assessing its physical consequences.

The paper is organized in the following way. In Section II we formulate the problem by
stating the assumptions, notation and scaling, leading to the basic set of non-dimensional
partial differential equations for the flow. In Section III the asymptotic expansions
of the various dependent variables are given and the equations are solved in each
order in the small parameter $\epsilon$, up to the second order. As already stated, the lowest order gives
the steady Shakura-Sunyaev solution and the higher orders provide analytically the
time dependent solution for a given initial disturbance. The most interesting effect---the \textit{transient growth}
 in the density and vertical velocity---occurs at the second order (Section III-D).
In Section IV and V, the solutions are discussed in some detail and we examine with the help
of graphic visualization some of their important physical properties
and finally, in Section VI, we summarize the results and their meaning.
Since the analytical procedure needed for obtaining the solutions, in various orders, is lengthy
and involved we defer some technical details to the Appendices.

\section{Formulation of the problem}
\label{equations}
\subsection{Notation, scaling and basic assumptions.}
The problem is formulated in cylindrical coordinates $(r,z,\phi)$ with $P$ and $\rho$ denoting the pressure and density functions.
The cylindrical components of the velocity are
$u, v$ and $r \Omega$, where $\Omega$ is the angular velocity. Additionally, $c_s$ is the sound
speed and $\eta$ the dynamic viscosity. The accretion flow is in the Newtonian gravitational field
of a central object with point mass  $M$.

The coordinate $r$ is scaled by its value at a typical point $r_\ast$ (we shall denote the scaling
dimensional variables by an asterisk) and the density by a typical
density value $\rho_\ast$. The {\em first  assumption} is a polytropic pressure-density relation $P=K\rho^{1+1/n}$,
with $n$ being the polytropic index and $K$ a constant.
Consequently the pressure is scaled by $P_\ast=K  \rho_\ast^{1+1/n}$. This also gives the
typical sound speed (squared) $ c_{s\ast}^2= d P_\ast/d \rho_\ast$. We shall scale the meridional
velocities with $c_{s \ast}$ and express the rotational angular velocity
in units of its Keplerian value at  $r_\ast$, so that $ \Omega_\ast^2=G M/ r_\ast^3$.
This then allows us to express the vertical coordinate scale, $h_\ast$, using the {\em second assumption},
that the flow is approximately in  vertical  equilibrium due to thermal pressure support
(that is, $h_\ast = c_{s\ast} /\Omega_\ast$). Our {\em third assumption} is that the
disk flow is cold, or, equivalently, that the azimuthal rotational velocity is highly supersonic. This is
equivalent to assuming very efficient cooling, by , e.g., radiative losses from the surface of the disk.
As a result, we find that $\epsilon \equiv  h_\ast/ r_\ast =  c_{s\ast}/( \Omega_\ast r_\ast)\ll 1$,
which happens to also measure the disk's thinness, as we alluded to in the Introduction.
This small number $\epsilon$ will thus appear in the non-dimensional equations characterizing the flow
and becomes our natural expansion parameter.

\subsection{Equations}

The non-dimensional polytropic relation is $P = \rho^{1+1/n}$ and we
define, for convenience, a function $W \equiv \int dP/\rho$. This gives supplementary polytropic relations, which
will ultimately allow to express $P, c_s$ and $W$ in terms of the density alone.
\beq
W  =n \frac{dP}{d\rho} =n c_s^2= (n+1)\rho^{1/n}~~~ {\rm because}~~~ c_s^2=(1+\frac{1}{n})\rho^{1/n}
\label{W}
\eeq
The full (time-dependent, non-axisymmetric) scaled hydrodynamic equations in cylindrical co-ordinates
read
\beqa
&&\epsilon \left\{ \partial_t u +   v \partial_z u
+ \Omega \partial_\phi u
\right\} + \epsilon^2 \{ u \partial_r u \}  =  \Omega^2r
- \frac{1}{r^2}\left(1 + \epsilon^2\frac{z^2}{r^2}\right)^{-\frac{3}{2}}
+ \epsilon \left(\frac{1}{\rho}\right)\left\{\partial_z\left(\eta \partial_z u \right)\right\}+
\nonumber \\
&& + \epsilon^2 \left(\frac{1}{\rho}\right)
\left\{ - \rho \partial_r W + \partial_z \left(\eta\partial_r v\right) -
\frac{2}{3} \partial_r \left(\eta \partial_z v \right)
+\partial_\phi \left(\eta\partial_r\Omega\right) - \frac{2 \eta}{r}\partial_\phi\Omega
-\frac{2}{3}\partial_r\left(\eta\partial_\phi \Omega\right)
\right\}+
\nonumber\\
&&~~~~~~~ + \epsilon^3 \left(\frac{1}{\rho}\right)
\left\{-\frac{2\eta u}{r^2} + \frac{2}{r}\partial_r
\left( \eta r \partial_r u \right)  - \frac{2}{3}\partial_r
\left[\frac{\eta}{r}\partial_r(ru)\right]
+\frac{1}{r^2} \partial_\phi\left(\eta \partial_\phi u \right)
\right\}
\label{horizontal}
\eeqa
\beqa
&& \partial_t \Omega + v \partial_z \Omega
+ \Omega \partial_\phi \Omega =
\left(\frac{1}{\rho}\right)\partial_z\left(\eta \partial_z\Omega\right)
-\epsilon \left\{\frac{u}{r^2} \partial_r\left(r^2\Omega\right) \right\}+
\nonumber \\
&&~~~~
+\epsilon^2 \left(\frac{1}{\rho r^2}\right)
\left\{
\frac{1}{r}\partial_r \left(\eta r^3 \partial_r \Omega \right)
-\rho \partial_\phi W +\frac{4}{3} \partial_\phi\left(\eta
\partial_\phi \Omega\right)+ \partial_z\left(\eta\partial_\phi v\right)
-\frac{2}{3}\partial_\phi\left(\eta\partial_z v\right)
\right\}+\nonumber\\
&&~~~~~~+ \epsilon^3 \left(\frac{1}{\rho r^2}\right)
\left\{
\frac{1}{r}\partial_r\left(r \eta \partial_\phi u \right) - \frac{2}{3} \partial_\phi
\left(\eta \partial_r u \right) + \frac{4}{3r}\partial_\phi \left(\eta u\right)
\right\}
\label{angular}
\eeqa
\beqa
&&  \partial_t v + v \partial_z v
+ \Omega \partial_\phi v
+\epsilon  \{ u \partial_r v \}  = - \partial_z W
- \frac{z}{r^3}\left(1 + \epsilon^2\frac{z^2}{r^2}\right)^{-\frac{3}{2}}
 +\frac{1}{\rho} \left[\frac{4}{3}\partial_z\left(\eta\partial_z v\right)
 + \partial_\phi\left(\eta \partial_z \Omega\right) - \frac{2}{3} \partial_z\left(\eta
 \partial_\phi \Omega\right)
 \right]+
\nonumber \\
&&~~~~~~~~~~~~~~~~~~~~~~~
+ \epsilon \left(\frac{1}{\rho r}\right) \left\{\partial_r\left(\eta r
\partial_z u \right)-\frac{2}{3}\partial_z\left[\eta\partial_r(ru)\right]\right\} +
 \epsilon^2 \left(\frac{1} {\rho r}\right)
\left\{ \partial_r \left(\eta r \partial_r v\right)
+ \frac{1}{r}\partial_\phi\left(\eta \partial_\phi v\right)
\right\}
\label{vertical}
\eeqa
\beq
\partial_t \rho + \partial_z \left(\rho v\right)
+\partial_\phi\left(\rho \Omega\right)
+ \epsilon \frac{1}{r}\partial_r\left(r \rho u \right)=0,
\label{cont}
\eeq
where (\ref{horizontal}-\ref{vertical}) are, respectively, the radial, azimuthal and vertical
momentum conservation equations, while (\ref{cont}) is the equation of mass conservation.
Note that this set of equations is very similar to the one in KK and UNRS.
The terms containing angle derivatives are missing in KK and UNRS, because of the
axisymmetry assumed there, while the terms containing time derivatives are missing in KK, who considered only
a steady state.\par
We impose a Lagrangian pressure condition on the vertical
surfaces (namely, that the pressure on the surface be zero).  We also require that the
stresses on the vertical surfaces be zero as well.  See Appendices C-D for details.
The radial boundary conditions require some discussion.
We will consider these equations for an extended ring, in which the internal radius is considerably
larger than the zero-torque radius of the disk $r_+$, that is, $r_* \gg r_+$ (see KK for a more extended discussion)
and the external radius
 is significantly smaller than the disk outer edge.
In this way we can avoid the treatment of inner and outer boundaries, which greatly complicates the problem,
presumably without changing the substantial result for the bulk
of the disk. Still it would be interesting to address in the future the dynamics at the inner edge region
in the case of black holes (using a different scaling and possibly matched asymptotic expansion)
and in the boundary layer in the case of neutron stars (by means of matched asymptotic expansion).
The outer boundary, at which the disk is fed by mass, depends strongly on the astrophysical system
in question and is of interest as well, but we defer also this question to future work.

\section{Asymptotic solutions}
\subsection{General}

The asymptotic expansion approach consists of expanding all functions in powers of $\epsilon$
(e.g., KK, UNRS) as follows
\beq
f(r,z,\phi,t)=f_0(r,z)+ \epsilon~\tilde f_1(r,z,\phi,t) + \epsilon^2~\tilde f_2(
r,z,\phi,t)...
\label{asymptoticexp}
\eeq
We postulate, similarly to UNRS, that the expansions of all the dependent variables are such
that the terms $\tilde f_j$ can be split into a spatial steady part $f_j$---the
steady base flow---and a time
dependent dynamical disturbance $f_j'$
\beq
\tilde f_j(r,z,\phi,t)=f_j(r,z,\phi)+f'_j(r,z,\phi,t)
\label{economy}
\eeq
\noindent
This form presupposes therefore that the time dependence is included only as an
additive function and  from the first order in $\epsilon$ and on.
\noindent We remark here at the outset that the viscosity, $\eta$, is not expanded.
Instead, we shall express it, using a standard prescription (as found in the Shakura-Sunyaev
solution \cite{SS}) by the lowest order
dependent variables (see section III C).

Without assuming anything additional we obtain what we call the {\em complete} set of equations in the first
three orders in $\epsilon$---see in Appends. \ref{compO0} - \ref{compO2}, respectively.
We then take the steady base flow to be the KK solution and so it is {\em axisymmetric}.
Although, this is very similar to the UNRS approach,
we shall not exclude a priori from the expansions terms which were
excluded by UNRS, who set them identically to zero, in accord with the work of KK.
These include the following variables:
$u_0,\tilde \Omega_1, ~v_0, ~\tilde v_1 , ~\tilde  \rho_1,$ and thus $~\tilde P_1~{\rm and}~\tilde W_1$.
Along our work we shall explicitly state
which assumptions of this kind can be "derived" and which are a matter of choice and thus are assumed.
Obviously, this work differs from UNRS also by allowing non-axisymmetric disturbances.
Note that
the vertical equilibrium at the zeroth order followed from the axisymmetry of the base flow---explicitly
of $\rho_0$.

Before turning to the equations and their solutions in the three lowest orders in $\epsilon$ we
assume, for the sake of simplicity,  that $n=3/2$  (as in KK). For our calculations we allow for
general values of $n$ and we agree with UNRS who find that the results are little influenced
 for reasonable $n$ values. For $n=3/2$, as we assume henceforth, we get
from the non-dimensional polytropic relation
and equation (\ref{W}) that
\beq
P_0=\rho_0^{5/3},\;W_0=\frac{5}{2}\rho_0^{2/3},\;c^2_{s0}=\frac{5}{3}\rho_0^{2/3}.
\label{W0}
\eeq
Finally, we remark that if $g(\rho)$ is a smooth function of the density alone,
and the density is written as $\rho=\rho_0 +\delta\rho$, where $\delta\rho$ is a small
perturbation atop $\rho_0$, then the perturbation in $g$, i.e., $\delta g$ is well
approximated by $\delta g = (\partial g/\partial \rho)_0 \delta \rho$. Thus
\beq
\delta W = \left(\frac{\partial W_0}{\partial \rho_0}\right) \delta\rho_0 =
\frac{5}{3}\rho_0^{-1/3} \delta\rho + \cdots
\label{Wpert}
\eeq
This result will be used in the orders $\epsilon$ and $\epsilon^2$  below.

\subsection{Order $\epsilon^0$}
\label{acompO0}
The complete equations at this order are simple because there is no time dependence
at this order. Since the steady base flow is also {\em axisymmetric}, it
further significantly simplifies the equations and gives a particularly compact equation set.
\begin{eqnarray}
\Omega _0^2 &=& \frac{1}{r^3}\label{ao01}\\
v_0 \rho _0 \frac{\partial \Omega _0}{\partial  z}
&=&
\frac{\partial }{\partial  z}\left( \eta  \frac{\partial \Omega _0}{\partial  z}\right)
\label{ao02} \\
v_0 \frac{\partial v_0}{\partial  z}
&=&-\frac{z}{r^3}-
\frac{\partial W_0} {\partial  z}+\frac{4}{3}\frac{1}{\rho _0}
\frac{\partial}{\partial z}\left( \eta \frac{ \partial v_0}{\partial  z}   \right)
\label{ao03}\\
 \frac{\partial (\rho _0 v_0)}{\partial  z} &=&0
 \label{ao04}
\end{eqnarray}

\subsubsection{Solution}
\label{subsubsec:sol1}
Equation (\ref{ao01}) guarantees the Keplerian form  $\Omega_0=r^{-3/2}$, which makes
equation (\ref{ao02}) trivial, and equation (\ref{ao04}) immediately gives
$$
\rho_0 v_0 = f(r).
$$
Since $v_0$ is an odd function in $z$ (and $\rho_0$ is even in $z$), we
get that $v_0=0$ identically and equation (\ref{ao03}) can be easily solved
(subject to the boundary condition $\rho_0 (r,h)=0$, where $h(r)$ is the height of the disk) to yield
$W_0(r,z)$ and thus $P_0(r,z)$ and $\rho_0(r,z)$. Thus the $\order 0$ solution is
identical to the KK solution at the same order,
\beqa
\Omega_0=r^{-3/2},~~~~~~~~~~~~~ v_0=0,~~~~~~~~~~~~~
\rho_0(r,z)=\left( \frac{h^2-z^2}{5 r^3} \right)^{3/2},
\label{ao0sol}
\eeqa
with $h=h(r)$ as given in KK and in UNRS,
\beq
\label{dimensionless_h}
{h(r) \over r}= \bar \lambda_0\left(1- \sqrt{{r_{+} \over r}}\right)^{1/6}
\hspace*{1.0cm} {\mbox{with}} \hspace*{1.0cm} \bar \lambda_0=\left[{\dot{M}
\over \alpha} \left({80 \over 3\pi}\sqrt{{5 \over 3}}\right)
\right]^{1/6}.
\eeq
\noindent Also
\begin{equation}
P_0(r,z)=\left( \frac{h^2-z^2}{5 r^3} \right)^{5/2},~~~~~~~~~~~~~W_0(r,z)={5 \over 2}
\left( \frac{h^2-z^2}{5 r^3} \right),
\end{equation}
following from (\ref{W0}).

\subsection{Order $\epsilon^1$}

The complete $\order{\epsilon}$ system, as given in Append. \ref{compO1} is significantly
simplified, when the axisymmetry of $\Omega_0$ and the fact that $v_0=0$, from the zero-order solution (\ref{ao0sol}),
are used.\\
Taking only the "unperturbed" parts gives rise to the following
time independent equation set
\beq
-2 r \Omega _0  \Omega _1 = \frac{1}{\rho_0}\frac{\partial}{\partial z}
\left(\eta \frac{\partial u_0}{\partial z}\right)\label{uao11}
\eeq
\beq
\frac{u_0}{r^2} \frac{\partial \left ( r^2 \Omega _0 \right)}{\partial r}
=\frac{1}{\rho_0}\frac{\partial}{\partial z}\left( \eta \frac{\partial \Omega _1}{\partial z} \right)\label{uao12}
\eeq
\beq
0= -\frac{\partial  W_1}{\partial z}
+\frac{4}{3\rho_0}\frac{\partial}{\partial z}\left(\eta \frac{\partial  v_1}{\partial z} \right)
-\frac{2}{3}\frac{1}{\rho_0 r}\frac{\partial}{\partial z}\left(\eta \frac{\partial (u_0 r)}{\partial r} \right)+
\frac{1}{\rho_0 r}\frac{\partial}{\partial r}\left(\eta r \frac{\partial u_0}{\partial z} \right)\label{uao13}
\eeq
\beq
\frac{1}{r} \frac{\partial \left( r \rho _0 u_0 \right)}{\partial r}+
\frac{ \partial \left( \rho _0 v_1 \right)}{\partial z}
=0\label{uao14}
\eeq
These equations are for the axisymmetric base flow (identical to KK; UNRS did not consider this order
and took all the contributions to be zero). In any case, subtracting this from the complete equation set
gives
\beq
-2 r \Omega _0  \Omega'_1 = 0\label{ao11}
\eeq
This equation guarantees that
$\Omega'_1=0$ and we can use the notation $\Omega_1$ for
$\tilde \Omega_1$, because it is time independent in this case. The resulting second equation gives no additional information (time-dependent, and see below)
and the remaining time-dependent equations to this order are
\beq
\drv{v'_1}{t} + \Omega_0\drv{v'_1}{\phi}=
-\frac{\partial W'_1}{\partial z}
+\frac{4}{3\rho_0}\frac{\partial}{\partial z}\left(\eta \frac{\partial v'_1}{\partial z} \right)
\label{ao13}
\eeq
\beq
\drv{\rho'_1}{t}+ \Omega_0 \frac{\partial
 \rho' _1}{\partial \phi}= -
\frac{ \partial \left( \rho _0  v'_1 \right)}{\partial z}
\label{ao14}
\eeq
We note the appearance of the operator
\beq
{\cal D}_{\phi t} \equiv \drv{}{t}+ \Omega_0 \drv{}{\phi},
\label{opdef}
\eeq
which we shall use henceforth, for economy of notation, remembering also that $\Omega_0(r)=r^{-3/2}$.
The subscript "$\phi t$" in the definition of this operator has the purpose of reminding that it contains both
the time and angular derivatives.

\subsubsection{Solution}

In the previous order the explicit form of the viscosity $\eta$ was not needed, but now
the situation is not as comfortable. Following Shakura and Sunyaev\cite{SS}, we posit the following form of the
$(\phi,r)$ viscous stress tensor component
\beq
|\tau_{\phi r}|=\alpha P_0,
\eeq
where $\alpha$ is an adimensional parameter.
This gives
\beq
\eta r \left|\frac{d\Omega_0}{dr}\right|=\alpha P_0~~~~\to~~~\eta=\frac{2}{3} \alpha r^{3/2} \rho_0^{5/3}
\label{etaexplic}
\eeq

Notice that this prescription is vertically dependent, i.e., the vertical distribution of  stress is assumed to follow the pressure distribution.
We now substitute this $\eta$ together with the polytropic relation,
$\Omega_0=r^{-3/2}$ and use $\partial_z{W_0}=-  z/{r^3}$
(from equation (\ref{ao03})) in equations (\ref{uao11}-\ref{uao12}), which become, after some
straightforward algebra

\begin{eqnarray}
\frac{2 \alpha}{3}\rho_0^{2/3}r^3 {\partial^2 u_0 \over \partial z^2}-\frac{2 \alpha}{3}z
{\partial u_0 \over \partial z}+(2r\Omega_1) &=& 0 \label{asol1}\\
\frac{2 \alpha}{3}\rho_0^{2/3}r^3 {\partial^2 (2r \Omega_1) \over \partial z^2}-\frac{2 \alpha}{3}z
{\partial (2r \Omega_1) \over \partial z}- u_0 &=& 0
\label{asol2}
\end{eqnarray}
Now, since we consider only  even  (in $z$) solutions for $u_0$ and $\Omega_1$, it
can be shown (see Appendix \ref{apdx0}) that the only bounded solutions are the trivial solutions:
\beq
u_0(r,z)=\Omega_1(r,z)=0.
\label{zeros}
\eeq
Equations (\ref{uao11}-\ref{uao12}) thus become trivial and equations (\ref{uao13}-\ref{uao14}) assume the form
\beq
0= -\frac{\partial  W_1}{\partial z}
+\frac{4}{3\rho_0}\frac{\partial}{\partial z}\left(\eta \frac{\partial  v_1}{\partial z} \right)
\label{uao13s}
\eeq
\beq
\frac{ \partial \left( \rho _0  v_1 \right)}{\partial z}=0
\label{uao14s}
\eeq
Considering equation (\ref{uao14s}) first, we get
$\rho_0 v_1 = f(r)$, but since $v_1$ is an odd
function of $z$ (i.e., it is zero at $z=0$), this implies $v_1=0$ identically.
Similarly, from equation (\ref{uao13s}) we get $ W_1=f(r)$, but at the disk vertical edge
(some large enough $z$) this function must be zero.
Consequently we will have, similarly to UNRS,
\beq
 v_1(r,z)= W_1(r,z)=\rho_1(r,z)= 0.
\label{iden0-2}
\eeq
We are thus left with the equations for the perturbations (\ref{ao13})-(\ref{ao14}).
UNRS omitted in their expansion the first order terms of the vertical velocity and density, as well as their time
dependent perturbations. Here we shall allow for these perturbations and therefore
will have to solve equations (\ref{ao13})-
(\ref{ao14}). These equations can be rewritten as
\beq
{\cal D}_{\phi t}~ v'_1 =
-{5\over 3}\frac{\partial}{\partial z}\left(\rho_0^{-1/3}  \rho'_1\right)
+\frac{4}{3\rho_0}\frac{\partial}{\partial z}\left(\eta \frac{\partial v'_1}{\partial z} \right)
\label{tao13}
\eeq
\beq
{\cal D}_{\phi t} ~\rho'_1= -
\frac{ \partial \left( \rho _0  v'_1 \right)}{\partial z},
\label{tao14}
\eeq
where we have used $W'_1 = (2/3) (W_0/\rho_0)\rho'_1 = (5/3) \rho_0^{-1/3} \rho'_1$ (see equation (\ref{Wpert})).
Applying again ${\cal D}_{\phi t}$ to (\ref{tao13}) and substituting (\ref{tao14}) we get the single equation
\beq
{\cal D}_{\phi t}^2 v'_1 - {4 \over 3 \rho_0}{\cal D}_{\phi t}\frac{\partial}{\partial z}\left(
\eta\frac{\partial v'_1}{\partial z}    \right)
-\frac{5}{3}\rho_0^{2/3} \frac{\partial^2}{\partial z^2} v'_1 -\frac{25}{6}\frac{\partial (\rho_0^{2/3})}{\partial z}
\frac{\partial}{\partial z} v'_1 - \frac{5}{2}\frac{\partial^2 (\rho_0^{2/3})}{\partial z^2} v'_1=0,
\label{tao15}
\eeq
where $\eta$ is given in terms of $\rho_0$ by equation (\ref{etaexplic}).
The above differential equation can be written as
\beq
{\cal L} v'_1 = 0,
\label{optau}
\eeq
where the form of the linear differential operator ${\cal L}$ can be easily inferred from the above.
This form becomes explicit using equation (\ref{ao0sol})
\beq
\rho_0(r,z)=(5r^3)^{-3/2} (h^2-z^2)^{3/2},
\label{nonsep}
\eeq
the viscosity  in terms of the coordinates (and constants)
\beq
\eta(r,z)=\frac{2}{3} \alpha r^{3/2} \rho_0^{5/3}=\frac{2}{3\times 5^{5/2}}\alpha r^{-6}(h^2-z^2)^{5/2},
\label{etaexp}
\eeq
and remembering that $h$ is a known function of $r$ (see KK, UNRS).

Equation (\ref{optau}) can be solved analytically.
At this purpose we introduce the similarity variable $T \equiv \Omega_0 t =t/r^{3/2}$ and the Ansatz $v'_1= \hat v_1(r,z, \phi) e^{s \, T} + \rm{c.c.}$, where $s$ is an eigenvalue.
Note that $v_1'$ must be periodic in $\phi$, with period $2 \pi$, and thus it can be Fourier expanded
and the Ansatz takes the form
\beq
v_1'(r,z, \phi, t)=\sum_{m=-\infty}^{\infty} \hat v_{1(m)}(r,\zeta) e^{sT + i m \phi} + {\rm c.c.},
\label{Fourier}
\eeq
where $\zeta \equiv z/h(r)$ is an another similarity variable and the lower index in parentheses denotes the
Fourier components.
Now, for each particular Fourier component, the Ansatz and the resulting eigenvalue equation are
like the ones discussed in UNRS (see equation (48) of that paper, which is for the case of
axisymmetry, i.e., $m=0$). The procedure is thus identical.

We summarize here the main result (which will also be used in the discussion of the next order).
The {\em ordinary} differential equation resulting from the substitution of the above
Ansatz, is known as the Gegenbauer (or hyper-spherical) equation and its solutions are known in terms
of combinations of the associated Legendre functions (also known as Gegenbauer polynomials), e.g., \cite{Math},
\cite{Abra}. These analytical expressions are also included in Wolfram's Mathematica 6 software,
which makes our calculations feasible. Applying the upper (and lower) boundary conditions gives
a condition on the eigenvalues. All the eigenvalues thus obtained have negative real parts (i.e., they decay in time).
For example, the fundamental mode for the Fourier component $m$ is
\beq
s_{(m)}^{\pm}= -\frac{4}{9}\alpha - i m \pm i \left|\frac{16}{81}\alpha^2-\frac{8}{3}\right|^{1/2}.
\label{eigen1}
\eeq
This is correct for any $\alpha \la 4$, other values of this parameter being un-physical for the
problem at hand.
So actually the fundamental mode solution is

\beq
v'_1(r,z,\phi ,t)= \sum_{-\infty}^{\infty} \hat v_{1(m)}(r,\zeta)e^{\mathit{i}  m \phi}
\left[S_+ e^{s_{(m)}^+ T + \mathit{i}  m \phi } +
S_- e^{s_{(m)}^- T }\right]+ \rm{c.c.},
\eeq
where $S_+$ and $S_-$ just are (integration) constants. Details can be found in Appendix \ref{apdxdecay}.

Clearly, the first order solutions found here
lead to solutions {\em exponentially} decaying in time . Since in this work we are interested to find transiently growing solutions,
we can reasonably
 make the choice of initial conditions such that $v'_1=\rho'_1=W'_1=P'_1=0$ at all times (as
it was assumed a priori in UNRS). In any case, the decaying solutions, even if not chosen to be zero
initially, become quickly negligible when the algebraically transiently growing solution appear (see next
order for a discussion). Moreover we stress that, whatever choices are made, they do not affect at all the solution at the next order (see
 \ref{U1}, \ref{U2}, \ref{U3} and \ref{U4}).

\subsection{Order $\epsilon^2$}

The second order equations are complicated (see Appendix \ref{compO2}), but taking only the "unperturbed"
parts (base flow), which are axisymmetric and steady, one gets the following  equation set:

\beqa
- 2r \Omega_0 \Omega_2
&=&-\drv{W_0}{r}+\frac{3z^2}{2 r^4}+\frac{1}{\rho_0}\opdrv{z}\left(\eta \drv{ u_1}{z}\right)
\label{KK1}\\
\frac{ u_1}{r^2}
\opdrv{r}\left(r^2 \Omega_0\right) &=& -\frac{1}{r^3\rho_0}\opdrv{r}\left(r^3\eta\drv{\Omega_0}{r}\right)
+\frac{1}{\rho_0}\opdrv{z}\left(\eta \drv{\Omega_2}{z}\right)\label{KK2}\\
0 =
 -\drv{ W_2}{z}+\frac{3 z^3}{2 r^5}&+&
\frac{4}{3 \rho_0}\opdrv{z}\left(\eta \drv{ v_2}{z} \right)
-\frac{2}{3 r\rho_0}\opdrv{z}\left[ \eta \drv{(r  u_1)}{r} \right]
+\frac{1}{r\rho_0}\opdrv{r}\left( r \eta \drv{  u_1}{z} \right)
\label{KK3}
\eeqa
\beq
0=\frac{1}{r}\opdrv{r}\left(r \rho_0 u_1 \right)+
\drv{(\rho_0 v_2)}{z}\label{KK4}
\eeq
The above equations (\ref{KK1}-\ref{KK4})  are identical
to the corresponding steady equation set of KK and UNRS (22-25).

After subtracting from the complete set we are left with the equations for the non-axisymmetric time-dependent perturbations
\beqa
\drv{u'_1}{t} + \Omega_0 \drv{u'_1}{\phi}&=&  2r \Omega_0 \Omega'_2
+\frac{1}{\rho_0}\opdrv{z}\left(\eta \drv {u'_1}{z}\right)
\label{U1}\\
\drv{\Omega'_2}{t}+ \Omega_0 \drv{\Omega'_2}{\phi}&=& -\frac{ u'_1}{r^2}
\opdrv{r}\left(r^2 \Omega_0\right)
+\frac{1}{\rho_0}\opdrv{z}\left(\eta \drv{ \Omega'_2}{z}\right)
\label{U2}\\
\drv{v'_2}{t} + \Omega_0\drv{ v'_2}{\phi} &=&
 -\drv{W'_2}{z}+
\frac{4}{3 \rho_0}\opdrv{z}\left(\eta \drv{ v'_2}{z} \right)
-\frac{2}{3\rho_0}\opdrv{z}\left(\eta \drv{ \Omega'_2}{\phi} \right)+\frac{1}{\rho_0}\opdrv{\phi}
\left(\eta\drv{\Omega'_2}{z} \right)-\nonumber\\
&&~~~~~~~~~ -\frac{2}{3 r\rho_0}\opdrv{z}\left[ \eta \drv{(r u'_1)}{r} \right]
+\frac{1}{r\rho_0}\opdrv{r}\left( r \eta \drv{ u'_1}{z} \right)
\label{U3}\\
\drv{\rho'_2}{t}+\Omega_0\drv{\rho'_2}{\phi} &=& -\frac{1}{r}\opdrv{r}\left(r \rho_0 u'_1 \right)-
\drv{(\rho_0 v'_2)}{z}+\rho_0\drv{\Omega'_2}{\phi}
\label{U4}.
\eeqa
Note that the LHS of all the four equations again contains the operator
${\cal D}_{\phi t} \equiv \partial_t + \Omega_0 \partial_\phi$.

\subsubsection{Reformulation in terms of linear differential operators}
We notice that
Equations (\ref{U1})-(\ref{U2}) dynamically decouple from Equations (\ref{U3})- (\ref{U4}), as in the axisymmetric
case, treated in UNRS.
The whole set of equations can thus be be reduced to
\begin{eqnarray}
{\cal P} u_1' &=&0\\
{\cal P} \Omega_2'&=&0\\
{\cal L} v_2'&=&[{\cal D}_{\phi t}{\cal F}+{\cal G}] u_1'+[{\cal D}_{\phi t} {\cal H}+{\cal J}]\Omega_2'
\label{inho},
\end{eqnarray}
The linear differential operators ${\cal P},{\cal L}, {\cal F}, {\cal G}$ are identical to those in UNRS,
in particular the operator ${\cal L}$ is the one inferred (in the previous order) from equation (\ref{tao15})
but here we have, in addition, the non-axisymmetric operators ${\cal J}$ and ${\cal H}$. The operators
are given, in terms of the known function $\rho_0(z)$, $\eta$ (a known function of $\rho_0$, r and constants)
as follows

\begin{eqnarray}
{\cal P} &\equiv& \left[ {\cal D}_{\phi t}-\frac{1}{\rho_0}\frac{\partial}{\partial z}\left(\eta \frac{\partial}{\partial z}
\right)\right]^2 + \Omega_0^2 \label{def_cal_P}\\
{\cal L}&\equiv& {\cal D}^2_{\phi t}-\frac{4}{3} \frac{1}{\rho_0} {\cal D}_{\phi t}
\frac{\partial}{\partial z}\left(\eta\frac{\partial}{\partial z}\right)-\frac{5}{3}\rho_0^{2/3}\frac{\partial^2}{\partial z^2}
-\frac{25}{6}\frac{\partial (\rho_0^{2/3})}{\partial z}\frac{\partial}{\partial z}
-\frac{5}{2} \frac{\partial^2 (\rho_0^{2/3})}{\partial z^2}\times \nonumber \\
{\cal F} &\equiv& -\frac{2}{3} \frac{1}{r \rho_0} \frac{\partial}{\partial z}\eta\frac{\partial}{\partial r}r +\frac{1}{r \rho_0}
\frac{\partial}{\partial r} r \eta \frac{\partial}{\partial z}, \label{def_cal_F}\\
{\cal G} &\equiv& \frac{2}{3}\frac{\partial}{\partial z} \frac{W_0}{r \rho_0}\frac{\partial}{\partial r}\rho_0 r, \label{def_cal_G}\\
{\cal H} &\equiv& -\frac{2}{3} \frac{1}{ \rho_0} \frac{\partial}{\partial z}\left(\eta\frac{\partial}{\partial \phi}
\right) +\frac{\eta}{\rho_0}
\frac{\partial^2}{\partial \phi \partial z}  \label{def_cal_H}\\
{\cal J} &\equiv&   \frac{5}{3}\frac{\partial}{\partial z}\left( \rho_0^{2/3}\frac{\partial}{\partial \phi}\right), \label{def_cal_J}
\end{eqnarray}
where we have used the expressions for the $n=3/2$ polytrope, and $\eta$ is as given in eq. (\ref{etaexp}).

\subsubsection{Solution}

We first find the eigenfunctions of the operator ${\cal P}$ (as in UNRS 50-51).
We then decide to consider just the fundamental modes to determine $u_1'$ and $\Omega_2'$ .
These can be substituted in the inhomogeneous equation (\ref{inho}) to get $v_2'$, in view of the the similarity between ${\cal F}$,
${\cal G}$ in UNRS and ${\cal H}$,${\cal J}$ here (the analogous solution was called in UNRS {\em driven acoustics}).
We shall discuss here only the case of $u_1'$ (the details can be found in Appendix \ref{appdxgrowth}), since the case of
 $\Omega_2^\prime$ is the same .

Mindful of the $2 \pi-$periodicity in $\phi$ we use the Ansatz
 \beq
 u_1'(r,z,\phi,t)= \sum_{m=-\infty}^{\infty} \hat u_{1(m)}(r,z) e^{p\,T + \mathit{i}  m \phi }
 + \rm{c.c.}
 \label{u1sol}
 \eeq
 and
\beq
 \Omega_2'(r,z,\phi,t)= \sum_{m=-\infty}^{\infty}
 \hat \Omega_{2(m)}(r,z) e^{p\,T + \mathit{i}  m \phi } + \rm{c.c.}.
 \label{om2sol}
\eeq
Note that the eigenvalue $p$ is different from $s$, while $T=t/r^{3/2}$ is defined as before.

The substitution of the Ansatz into ${\cal P} u_1^\prime=0$ gives rise, for each $m$, to a differential equation
that is somewhat more complicated than the Gegenbauer equation we had in the first
order:
\beq
\left\{\left[ \frac{1}{5}\alpha(1-\zeta^2)\frac{\partial^2}{\partial\zeta^2}-\alpha\zeta\frac{\partial}
{\partial\zeta} -\frac{3}{2}( p + \mathit{i} m)\right]^2 +\frac{9}{4}\right\} \hat u_{1(m)}(r,\zeta)=0.
\label{Frob}
\eeq
Still the equation can be tackled analytically, using
the series expansion (Frobenius) method. As shown in Appendix \ref{appdxgrowth}
the spatial part of the fundamental mode ($k=0$) has a rather simple structure
\begin{equation}
\hat u^{(0)}_{1(m)}(r,\zeta)  = A(r)\left(\zeta^2 - \frac{1}{6} \right),
\label{u_solution}
\end{equation}
and, similarly
\begin{equation}
\hat \Omega^{(0)}_{2(m)}(r,\zeta)  = C(r)\left(\zeta^2 - \frac{1}{6} \right).
\label{om_solution}
\end{equation}
The appropriate eigenvalue $p$, following from the surface boundary conditions, can take two values
\begin{equation}
 p_{(m)}^{\pm}= - \frac{8}{5}\alpha - i m \pm i .
\label{eigenvalue_2}
\end{equation}

So equation (\ref{u1sol}) should be replaced by
\beq
u'_1(r,z,\phi ,t)= \sum_{-\infty}^{\infty} \hat u_{1(m)}(r,\zeta) e^{\mathit{i}  m \phi}
\left[ P_+ e^{p_{(m)}^+ T} + P_- e^{p_{(m)}^- T} \right] +\rm{c.c.}, \label{u1sol_mod}
\eeq
where $P_{\pm}$ are constants. The same holds for $\Omega_2'$
 We note that the homogeneous part of  the linear inhomogeneous equation (\ref{inho}) is
identical to the equation considered in the first order (i.e., involving the operator ${\cal L}$) and that the
RHS  (the inhomogeneous part) can be found using the solutions for $u_1'$ and $\Omega_2'$ that we just discussed.
As is well known, the general solution of the inhomogeneous equation  ($v_2'$) is the sum of the general solution to the homogeneous equation
(denoted by $v_h'$ ) and a particular solution to the inhomogeneous one ($v_p'$)

\begin{equation}
v_2'=v_h'+v_p'.
\end{equation}

We already know that $v_h'$ is an exponentially time decaying function.
Since  $v_2'$ has to be $2 \pi-$periodic in $\phi$ we make a Fourier expansion. Due to the linearity of the operators,
$v_h'$,$v_p'$,$u_1'$ and $\Omega_2'$ must have the same $\phi-$dependence.
The structure of the operators ${\cal L}$,${\cal F}$,${\cal G}$,${\cal H}$ and ${\cal J}$ points to
a particular solution in the form
\begin{equation}
v_p' =
\sum_{k=0}^{\infty}\sum_{m=-\infty}^{\infty}
e^{\mathit{i} m\phi}\left[\hat v^{(k)+}_{(m)}(r,\zeta)e^{p_{(m)}^+ T}
+ \hat V^{(k)+}_{(m)}(r,\zeta) T e^{p_{(m)}^+ T}
+\hat v^{(k)-}_{(m)}(r,\zeta)e^{p_{(m)}^- T }
+ \hat V^{(k)-}_{(m)}(r,\zeta) T e^{p_{(m)}^- T} \right] \ \ +  {\rm c.c.},
\label{sol_v}
\end{equation}
with $T$ and $\zeta$ as before and where we include, for generality, also the
overtones $k\neq 0$.
Inserting the above expression for $v_p'$ in equation (\ref{inho}) and substituting on the RHS
the solutions based on equations (\ref{u1sol}), (\ref{om2sol}), (\ref{u_solution}) and (\ref{om_solution}),
we are led to  a rather long and involved expression (see Appendix \ref{appdxgrowth}, equation \ref{eq_app}).
The fundamental mode
($k=0$ and dropping the vertical mode superscript)
has the following form (the superscript $\pm$ corresponds to the two eigenvalues $p_{(m)}^\pm$)
\begin{equation}
\hat v^{\pm}_{(m)} = a_{3(m)}^{\pm}(r)  \zeta^3 + a_{1(m)}^{\pm}(r) \zeta,\qquad
\hat V^{\pm}_{(m)}= b_{3(m)}^{\pm} (r) \zeta^3 + b_{1(m)}^{\pm}(r) \zeta.
\label{fundamental_v}
\end{equation}
By setting to zero the coefficients of $t$ and each power of $\zeta$ we get $ a_{1(m)}^{\pm}(r)$,
$a_{3(m)}^{\pm}(r)$,  $b_{1(m)}^{\pm}(r) $ and $b_{3(m)}^{\pm}(r) $.

One may follow a similar procedure for the density perturbation, using
\begin{equation}
\rho_2'=\rho_h'+\rho_p'.
\end{equation}

\noindent Following the argument outlined in Appendix D, we get that $\rho_h'\equiv 0$.
The Ansatz
\begin{equation}
\rho_p' =
\sum_{k=0}^{\infty}\sum_{m=-\infty}^{\infty} e^{\mathit{i}m\phi}
\left[\hat \rho^{(k)+}_{(m)}(r,\zeta)e^{p_{(m)}^+ T}
+ \hat R^{(k)+}_{(m)}(r,\zeta) T e^{p_{(m)}^+ T} +\hat \rho^{(k)-}_{(m)}(r,\zeta)e^{p_{(m)}^- T}
+ \hat R^{(k)-}_{(m)}(r,\zeta) T e^{p_{(m)}^- T}\right] \ \ +  {\rm c.c.}
\end{equation}
\noindent can be substituted in equation
(\ref{U4}) to find for the fundamental mode
\beqa
\hat \rho^{\pm}_{(m)} &=& (1-\zeta)^{\frac{1}{2}}
\left[c_{4(m)}^{\pm} \zeta^4 + c_{2(m)}^{\pm} (r) \zeta^2 +c_{0(m)}^{\pm}(r)\right],\nonumber\\
\hat R^{\pm}_{(m)} &=& (1-\zeta)^{\frac{1}{2}} \left[d_{4(m)}^{\pm} (r) \zeta^4 +
d_{2(m)}^{\pm}(r)\zeta^2 + d_{0(m)}^{\pm}(r)\right].
\label{fundamental_rho}
\eeqa

The functions  $a_{i(m)}^{\pm}$,$b_{i(m)}^{\pm}$, $c_{i(m)}^{\pm}$ and $d_{i(m)}^{\pm}$ are known analytical, generally complex, functions of $r$ and depend
also on the parameter $\alpha$.
Being very complicated expressions, we handle them using the symbolic tools in Wolfram's Mathematica 6, but we do not write them in the
paper explicitly. Finally we can rewrite, in a more compact way, the fundamental
driven (by the horizontal velocity perturbations) vertical acoustic modes as

\begin{equation}
\left[
\begin{array}{c}
\hat v^{\pm}_{(m)} \\
\hat V^{\pm}_{(m)}
\end{array}
\right]
=
\left[
\begin{array}{c}
a_{1(m)}^{\pm} \\
b_{1(m)}^{\pm}
\end{array}
\right]\zeta
+
\left[
\begin{array}{c}
a_{3(m)}^{\pm} \\
b_{3(m)}^{\pm}
\end{array}
\right]\zeta^3,
\end{equation}

and

\begin{equation}
\left[
\begin{array}{c}
\hat \rho^{\pm}_{(m)} \\
\hat R^{\pm}_{(m)}
\end{array}
\right]
=
\bigl(1-\zeta^2\bigr)^{\frac{1}{2}}
\left\{
\left[
\begin{array}{c}
c_{0(m)}^{\pm} \\
d_{0(m)}^{\pm}
\end{array}
\right]
+
\left[
\begin{array}{c}
c_{2(m)}^{\pm} \\
d_{2(m)}^{\pm}
\end{array}
\right]\zeta^2
+
\left[
\begin{array}{c}
c_{4(m)}^{\pm} \\
d_{4(m)}^{\pm}
\end{array}
\right]\zeta^4
\right\}.
\label{radial_functions}
\end{equation}

\section{TEMPORAL EVOLUTION OF THE PERTURBATION ENERGY}

Armed with analytical expressions for the dynamical variables of the disk, that are
solutions of an IVP in which
the initial conditions are all small perturbations on a steady base flow,
we may now consider possible physical implications of our analysis.

The solutions we found point to an exponential time decay of most variables, but there is also a
 transient fast algebraic growth (before an ultimate, rather slow, exponential decay) of
disturbances of order $\epsilon^2$ in two variables---the vertical velocity $v_p'$ and density $\rho_p'$.
These quantities can be naturally interpreted as sound waves, which were called in UNRS {\em driven acoustics}.
UNRS study was axisymmetric, while here we have allowed for azimuthal dependence. Exploiting the obvious
$2 \pi$ periodicity in $\phi$, however, we took recourse to Fourier series expansions in this angle
variable and thus were able to consider separately the various Fourier modes.

As in UNRS, we perceive the acoustic energy as a relevant variable to follow, but here we consider the
time dependence of the energy contained in various Fourier modes as well as the effect
of different values of $\alpha$ and of the (free) radial form of the initial conditions.
Following Rayleigh's book \cite{JWSR} (Chapter XI) we define the
acoustic energy density as the sum of the kinetic and potential energy densities.
The potential term is the work gained/lost during expansion/compression and can be easily found using $P'dV$ and
$P'=\rho_0 c_{s0}^2 \rho'$. Hence, the acoustic energy volume-density in the $m$-th Fourier component is
\beq
\varepsilon_{(m)}(r,z,\phi,t,\phi;\alpha) = \frac{1}{2}\rho_0  \tilde v_{(m)}^2 + \frac{1}{2}\frac{c_{s0}^2
\tilde \rho_{(m)}^2}{\rho_0}.
\label{eq:en}
\eeq
The total acoustic energy volume-density is
\beq
\varepsilon(r,z,\phi,t,\phi;\alpha) = \frac{1}{2}\rho_0  \tilde v^2 + \frac{1}{2}\frac{c_{s0}^2
\tilde\rho^2}{\rho_0},
\eeq
and we shall consider the fundamental $k=0$ mode only.

The zeroth order vertical velocity is zero, as well as $v_1(r,z,\phi)$ and  $\rho_1(r,z,\phi)$.
Thus remembering that
 $\tilde f=f(r,z,\phi)+f'(r,z,\phi,t)$, $v_2'=v_h'+v_p'$ and  $\rho_2'=\rho_p'$, we get
\beq
\tilde v= \epsilon v_1'+ \epsilon^2 (v_2+v_h'+v_p'), \qquad
\tilde \rho= \epsilon \rho_1'+ \epsilon^2 (\rho_2+\rho_p'),
\eeq
where (we repeat all solutions here, for completeness)

\begin{eqnarray}
v_1' &=& \sum_{m=-\infty}^{\infty}e^{\mathit{i}  m \phi} \hat v_{1(m)}(r,\zeta) \left[ S_+ e^{s_{(m)}^+ T  }+
S_-  e^{s_{(m)}^-T}\right]+\rm{c.c.}\nonumber\\
v_h' &=& \sum_{m=-\infty}^{\infty}e^{\mathit{i}  m \phi} \hat v_{h(m)}( r, \zeta) \left[ S_+^h e^{s_{(m)}^+ T  }+
S_-^h  e^{s_{(m)}^-T}\right]+ \rm{c.c.}\nonumber\\
v_p' &=& \sum_{m=-\infty}^{\infty} e^{\mathit{i}  m \phi}\left[\hat v^{+}_{(m)}(r,\zeta)e^{p_{(m)}^+ T}+
\hat V^{+}_{(m)}(r,\zeta) T e^{p_{(m)}^+ T+}+\hat v^{-}_{(m)}(r,\zeta)e^{p_{(m)}^- T}
+ \hat V^{-}_{(m)}(r,\zeta) T e^{p_{(m)}^-T}\right] +  {\rm c.c.}\nonumber\\
\rho_1' &=& \sum_{m=-\infty}^{\infty} e^{\mathit{i}  m \phi} \hat \rho_{1(m)}(r,\zeta) \left[ Z_+ e^{s_{(m)}^+ T  }+
Z_-  e^{s_{(m)}^- T }\right]+ \rm{c.c.}\\
%
%
\rho_p' &=& \sum_{m=-\infty}^{\infty}e^{\mathit{i}m\phi}\left[\hat \rho_{(m)}^{+}(r,\zeta)e^{p_{(m)}^+ T}+ \hat R^{+}_{(m)}(r,\zeta)
T e^{p_{(m)}^+ T}+
\hat \rho_{(m)}^{-}(r,\zeta)e^{p_{(m)}^- T}+ \hat R^{-}_{(m)}(r,\zeta) T e^{p_{(m)}^- T}
\right] +  {\rm c.c.}\nonumber,
\label{all_solutions}
\end{eqnarray}
where $S_\pm, S^h_\pm, Z_\pm$ are (integration) constants.
and where
\beq
T=\Omega_0 t=t/r^{3/2}, \qquad
s_{(m)}^\pm= -\frac{4}{9}\alpha - i m \pm \mathit{i} \left|
\left(\frac{16}{81}\alpha^2 -
\frac{8}{3}
\right) \right|^{1/2},  \qquad
p_{(m)}^\pm= -\frac{8}{5}\alpha - i m \pm \mathit{i}.
\eeq
We recall that $\hat v_{1(m)}$, $\hat v_{(h)m}$  are the appropriate Gegenbauer polynomials, while
$\hat v^{\pm}_{(m)}$, $\hat V^{\pm}_{(m)}$ are odd polynomials of $\zeta$ and $\hat \rho^{\pm}_{(m)}$, $\hat R^{\pm}_{(m)}$ and $\hat \rho_{1(m)}$
are even polynomials of $\zeta$.
The terms $v_1'$ and $v_h'$ exponentially decay while $v_p' \sim T e^{p T}$ grows linearly, so that
the algebraically growing  terms (like $\sim T$) dominate the particular solution.
Moreover, when its eventual decay takes over,
the former decaying terms are negligible (for the relevant very small values of $\alpha$).
So, for simplicity we have decided to take the initial conditions such
that $v_1'$ and $v_h'$ are zero and thus remain so for all times.
The same reasoning is valid for $\tilde \rho$.
Therefore when we calculate the acoustic energy we can take just
$\tilde v=v_p'$ and $\tilde \rho=\rho_p'$.

In order to study the {\em temporal} behavior of the acoustic energy, we define two integral quantities
 the radius-dependent---i.e, averaged over a ring---energy per unit area
of the disk ${\cal E}_r$ and the total acoustic energy of the fluctuation $E_a$

\beq
{\cal E}_r(r,t;\alpha,m) \equiv \int_{-h(r)}^{h(r)}\int_{0}^{2 \pi}{{\cal E}(r,z,t,\phi;\alpha,m)}dz d\phi, \qquad
E_a(t;\alpha,m) \equiv \int_{r_{\rm min}}^{r_{\rm max}}{{\cal E}_r \ r \ dr}.
\label{integrals}
\eeq

\begin{figure}
\begin{center}
\leavevmode \epsfysize=8.cm
\epsfbox{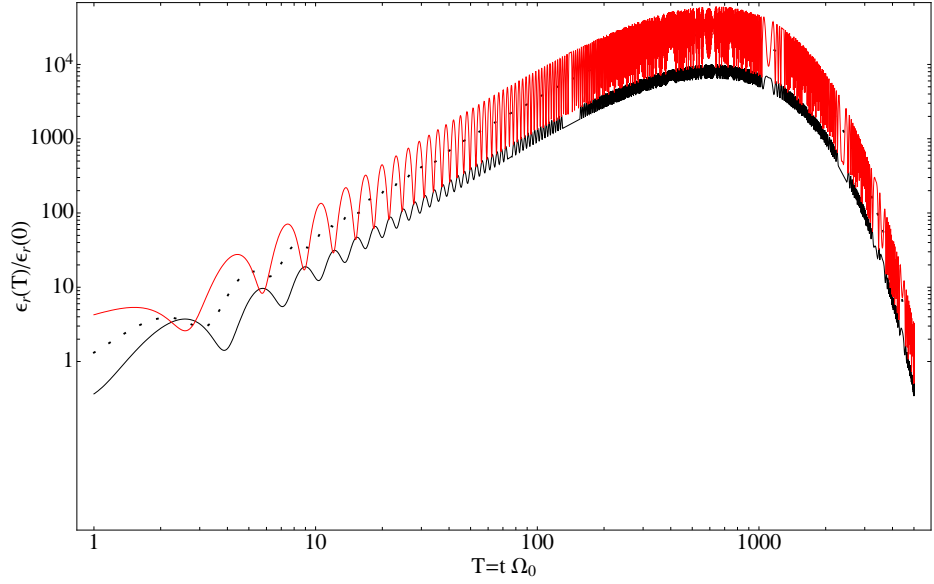}
\end{center}
\caption{{\small
Time evolution of the surface-density of the acoustic energy in the fundamental $k=0$ mode
at $r=1$, $\alpha=0.001$ and $A(r) =C(r)= e^{i\pi/4}$.  The three analytical curves are for
$m = 0$ (solid lower black line), $m=1$ (black dotted line), $m=10$ (solid upper red line) and are shown in a log-log plot.
${\cal E}_r(T)$ is scaled to its corresponding value at $T=0$. Notice that  ${\cal E}_r$ is
modulated by fast oscillations.
}}
\label{fig:enr}
\end{figure}

\subsection{The azimuthally averaged energy surface density}

\begin{figure}
\begin{center}
\leavevmode \epsfysize=8.cm
\epsfbox{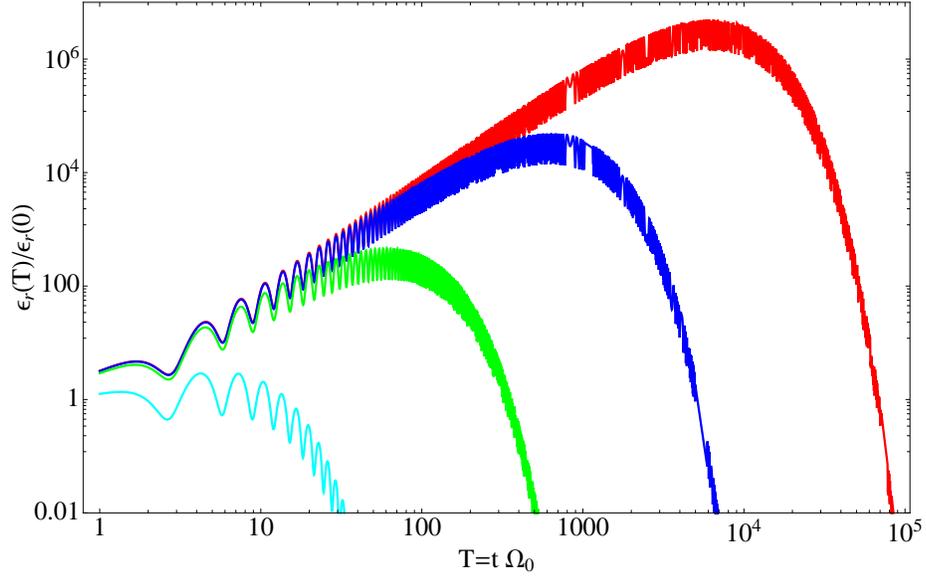}
\end{center}
\caption{{\small
Same as in Figure \ref{fig:enr}, but for a single Fourier component ($m=3$) and
different values of $\alpha$ . Four curves, for
$\alpha=0.0001,0.001,0.01,0.1$ (from the top to the bottom), are shown.
}}
\label{fig:enalpha}
\end{figure}

We obtain ${\cal E}_r$ for a particular Fourier mode $m$ in the form

\begin{equation}
{\cal E}_r(r,T;\alpha,m) = e^{-\frac{16}{5}  \alpha T} F(r;\alpha,m,\cos{2 T},\sin{2 T}),
\end{equation}

where the function $F$ is known analytically, but the expression is extremely long and we shall
not write it out here explicitly. The derivation is outlined in Appendix F.
Instead, we display the results graphically.
\begin{figure}
\begin{center}
\leavevmode \epsfysize=8.cm
\epsfbox{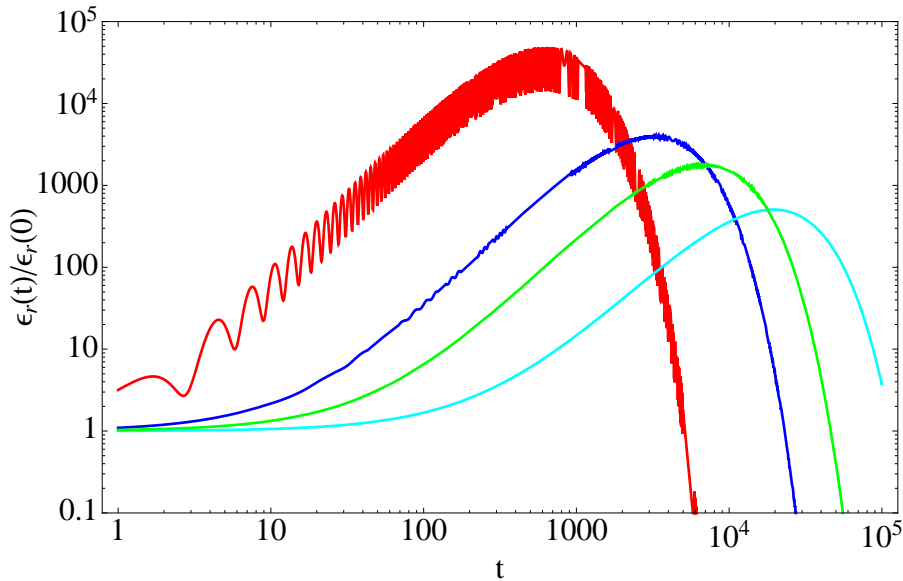}
\end{center}
\caption{{\small
Same as in Figure \ref{fig:enalpha}, but for one fixed
$\alpha=0.001$ and at different radial positions.  Four curves for
$r=1,3,5,10$ (from the top to the bottom) are shown.
}}
\label{fig:enradius}
\end{figure}

In the first three figures the temporal behavior of the normalized acoustic energy
surface-density contained in a particular Fourier component,
${\cal E}_r(r,t;m,\alpha)/{\cal E}_r(r,0;m,\alpha)$, is shown. This quantity is shown
as a function of the similarity variable $T$, defined before and for
the fundamental $k=0$ mode, in a log-log plot. In all the figures the free radial
functions are set to $A(r) =C(r)= e^{i\pi/4}$, which means that the initial perturbation
is taken, for simplicity, to be $r$-independent.

In Figure \ref{fig:enr}, $\alpha=0.001$ and $r=1$ are fixed and the various curves are for
different Fourier components. The most significant feature of the behavior displayed in this
figure (as well as in all subsequent ones) is the strong transient growth (by orders of magnitude) of the perturbation
energy, which typically takes place for a rather significant time.
The non-axisymmetric modes display maximal growth. All the mode have fast  oscillations, atop the much slower, secular growth.
From this Figure it appears that the maximum is attained at $t\approx 1000$ (since $r=1$,
we have here $t=T$).

Figure \ref{fig:enalpha} shows the dependence of the transient growth on the value of the $\alpha$
parameter, for the single Fourier component $m=3$ (other modes display a similar behavior).
The radius is fixed, as before, at $r=1$. We clearly see that the lower is the value of $\alpha$, the higher is the maximum,
and the later it occurs. While $\alpha=0.1$ does not give rise to any growth at all, for $\alpha=10^{-4}$
the growth is enormous---by a factor of a few times $10^5$. Since the growth occurs when taking into account
the $\epsilon^2$ terms of our expansion and the energy is composed of squares of these terms,
the validity of the expansion is marginal for such a growth (assuming $\epsilon \sim 10^{-2}$).
However for $\alpha \sim 10^{-3}$ (close to a "realistic" value, as found numerically
in the sub-critical hydrodynamic transition \cite{LL} ,
 the growth is somewhat less than $10^4$ and the asymptotic expansion reasonably holds.

In Figure \ref{fig:enradius} $\alpha=0.001$ and the Fourier component $m=3$ are fixed, while we show
the behavior  at different radii (from the top to the bottom $r=1,3,5,10$). The growth is higher and it reaches its maximum
earlier at smaller radii: clearly at large radii the perturbation becomes negligible.\\

Before moving on to the description of the behavior of the total acoustic energy, we would like
to remark that although our choice of the radial functions $A(r)$ and $C(r)$ may seem non-physical,
because it is difficult to imagine an $r$-independent perturbation,  it is still meaningful.
Indeed, since so far we have dealt with the angle averaged $r$-dependent surface density of the energy,
it is obvious what to expect when an $r$-localized perturbation is considered
instead. The results can be simply scaled, depending on the relative value of the perturbation
at the particular radius where they are sought (e.g., see Figure \ref{fig:enradius}). In the next subsection,
when we shall consider
the total (also $r$-integrated) energy in an extended ring, we will use perturbations that are $r$-localized
(in the form of a Gaussian).

\subsection{The total acoustic energy in a ring}

The total acoustic energy of the perturbation is evaluated as an integral over $r$
of the afore considered function
\beq
E_a(t;\alpha,m) \equiv \int_{r_{\rm min}}^{r_{\rm max}}{{\cal E}_r(r,t;\alpha,m) \ r \ dr}.
\eeq

The ring we consider is between $r_{\rm min}=1$ and $r_{\rm max}=10$, far from the inner and
 outer edge of the disk.


For simplicity, the radial functions of the perturbation will be taken in the form of Gaussians equal to each
other

\beq
A(r) =C(r)= e^{i\pi/4} e^{-(r-r_0)^2/\Delta^2},
\eeq
so that $r_0$ is the center of the perturbation and $\Delta$ its width.\\

In principle, our expansion is valid up to a time of the order of $\epsilon^{-2}$.
However in making the choice of $r_0$ and $\Delta$ we should take into account the fact
that, in general, the perturbation may propagate with the speed of sound. While the boundary
conditions on $z$ have been chosen in a physically sound way, we do not have specified any $r$-
boundary conditions (this being also the property of the KK and UNRS solutions). Thus, if our
ring is determined, we should not allow any wave to reach these non-physically natural boundaries,
so as not to create spurious effects (e.g., reflections that may artificially reinforce the perturbations).
Therefore the validity of our results should be limited in time.
We shall now estimate the time $\tau$ during which our result is valid. Let $\delta r$ be the
distance from $r_0$ to the nearest ring edge: then in our units an estimate
of the time for a sound wave to reach that edge is given by
\beq
\tau \sim \frac{\delta r}{c_s} \sim \frac{\delta r}{\Omega_0(r_0)h(r_0)}= \epsilon_0^{-1}\
\sqrt{r_0}\ \delta r,
\eeq
where $\epsilon_0= h(r_0)/r_0$. Our steady-solution, far enough from the zero torque radius, has
$h\propto r$ (see the previous section and KK) and thus we may substitute $\epsilon_0=\epsilon$.

For example, if we take $r_0=2$ and the inner edge is at $r=1$, we have $\delta r$=1 and so
$\tau \sim \sqrt{2}/\epsilon$ . Hence, for $\epsilon=10^{-2}$, the result should be reliable for only a little over
$100$ time units. The width of the perturbation further limits this. A better choice would be
to place the perturbation at the center of the ring, i.e., $r_0=5.5$: then we would have
$\tau \sim 4.5 \sqrt{5.5}/\epsilon$, which would give a validity of $\sim 1000$ time units.

\begin{figure}
\plottwo{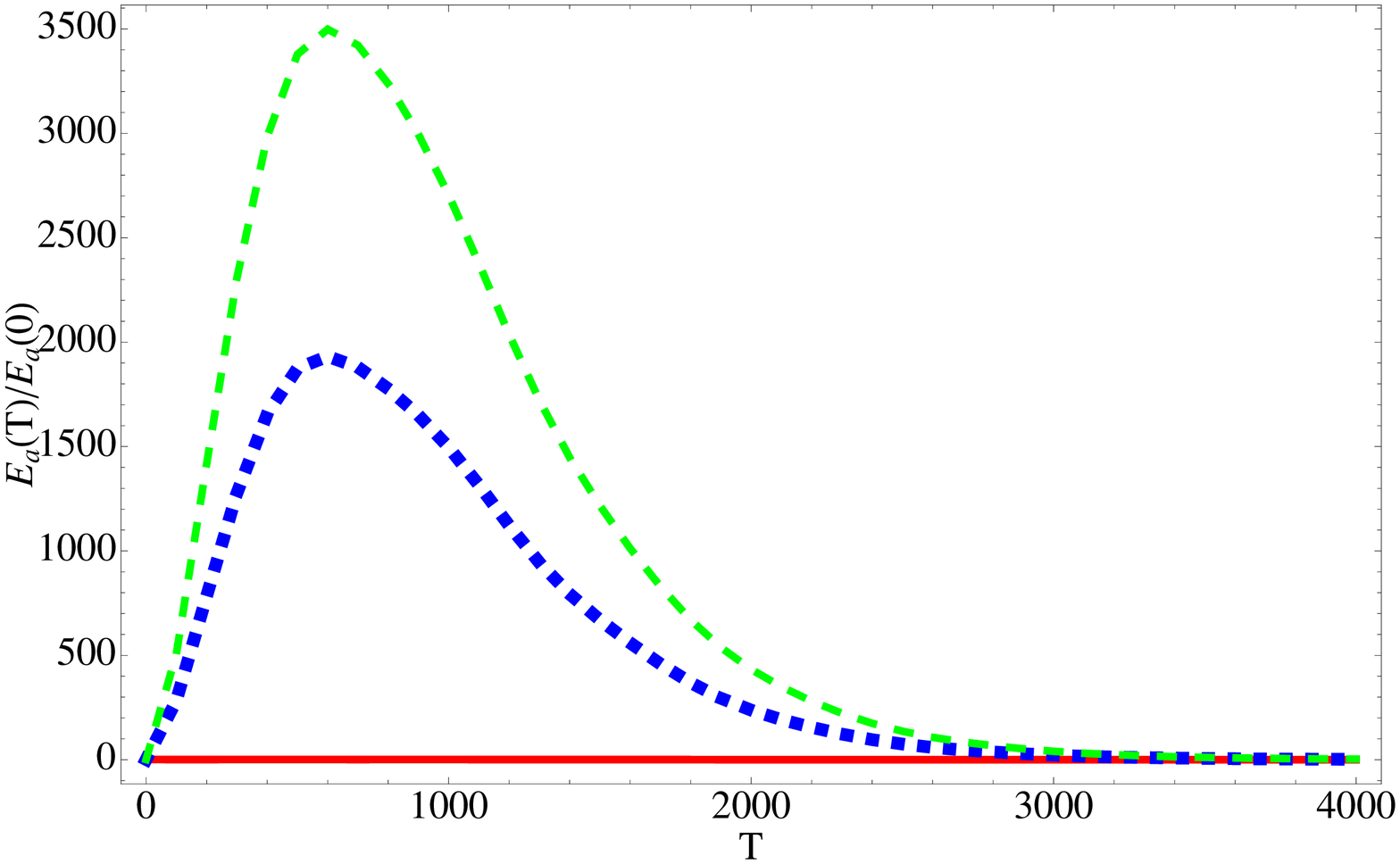}{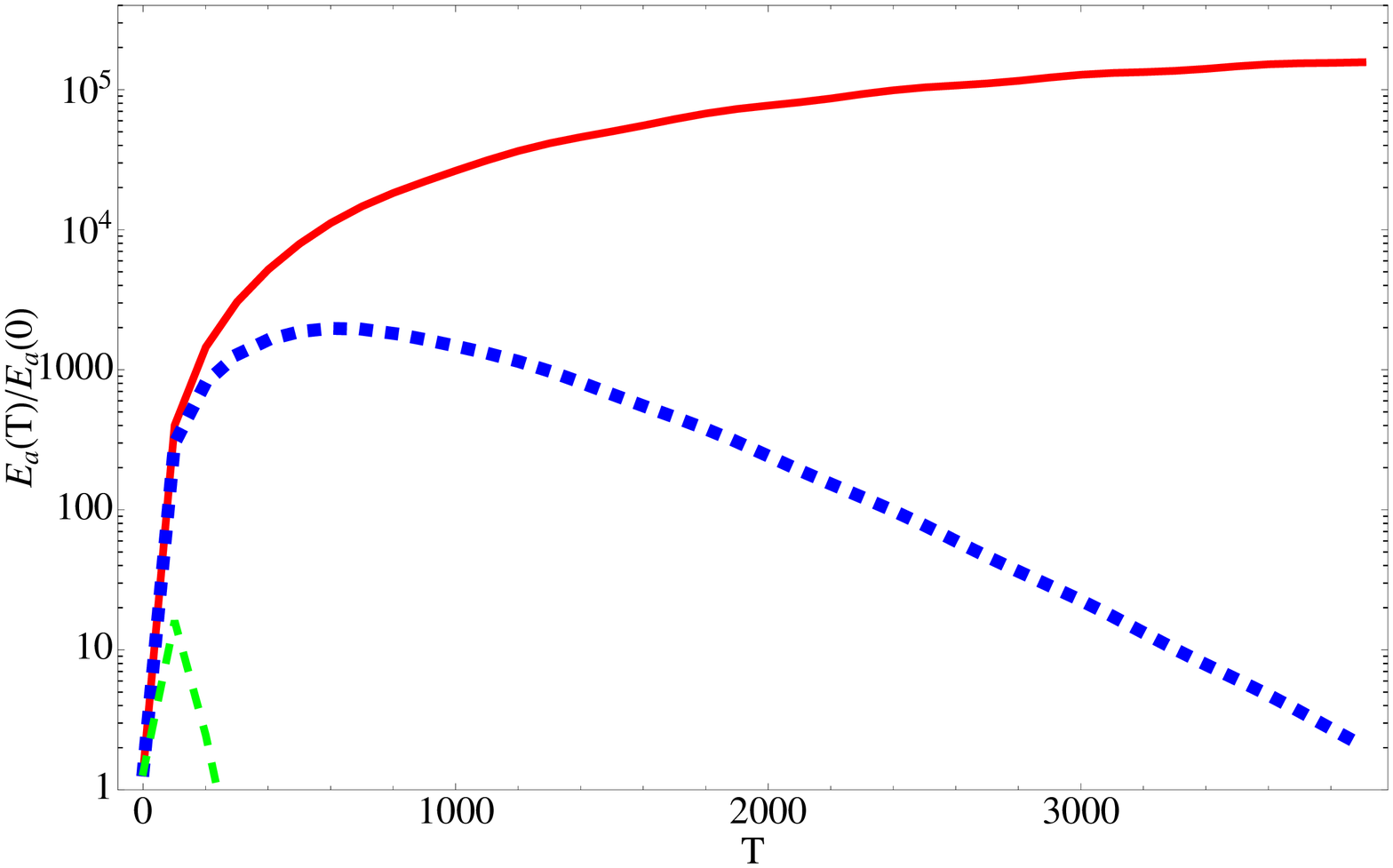}
\caption{{\small Log-log plot of the total acoustic energy in a ring.
The left panel is for different modes ($m=1,3,10$ from the bottom to the top). The right panel is for $m=3$ and
different viscosity ($\alpha=0.0001,0.001,0.001$ from the bottom to the top). For details see text.}}
\label{fig:enint}
\end{figure}

In Figure \ref{fig:enint} we take $r_0=4, r_{\rm max}=10$ , $\Delta=\sqrt{2}$ and
plot the evolution of the normalized energy in a ring, $E_a(T)/E_a(0)$.
The left panel is for different modes and it shows that the relative growth is higher for higher $m$. As expected,
the transient growth is more
pronounced for smaller viscosity (right panel).

\section{Surface density spatio-temporal behavior}

The formalism developed in UNRS and this paper can be used to follow the time evolution
(and, in particular, the transient growth) of various small perturbations, which are included
in the initial conditions of the appropriate IVP. We have already shown the copious transient
growth of the acoustic surface energy (averaged over the azimuthal angle) as well as of the total acoustic
energy in a finite ring of the disk. As suggestive as these results may be, for the possible
disk energetics, they do not contain explicit {\em spatio}-temporal dynamical information. The
possibilities to gain the latter are rather abundant, and it would be outside the scope of a single
paper to examine a great many of them, in detail. Thus, we have decided to conclude here, by calculating and
presenting just one of the important dynamical variables---the surface density.
Other information that can be extracted from our three-dimensional analytical solution, found in this paper,
will be presented in later works.

\subsection*{Expression for the leading terms}
In Section IV we have already estimated that the algebraically growing terms will be the dominant ones.
We shall repeat now this argument applied to the density.
Focusing on the fundamental vertical mode $k=0$ (thus dropping the superscripts) and taking only {\em one},
the $m$-th, say, Fourier component, we get for the
particular solution for the density perturbation, appearing in second order in $\epsilon$, the following
expression, resulting from equation (\ref{all_solutions}).
\beq
\rho_p' = \left[\hat \rho_{(m)}^{+}(z,r)e^{p_{(m)}^+ T+\mathit{i}m\phi}+\hat R^{+}_{(m)}(r,\zeta) T e^{p_{(m)}^+ T +\mathit{i} m \phi}
\right]+\left[\hat \rho_{(m)}^{-}(z,r)e^{p_{(m)}^- T+\mathit{i}m\phi}+ \hat R^{-}_{(m)}(r,\zeta) T e^{p_{(m)}^- T +\mathit{i} m \phi}
\right] +  {\rm c.c.},
\eeq
where $T\equiv t r^{3/2}$, $\zeta \equiv z/h(r)$ and $p_{(m)}^\pm=-8\alpha/5 -i (m \mp 1)$.

The full asymptotic series for the density includes, however, more terms and is rewritten
here, up to the second order in $\epsilon$
\beq
\rho(r,\zeta,\phi,t) = \rho_0 (r,\zeta) + \epsilon \left[ \rho_1(r,\zeta, \phi) + \rho_1'(r,\zeta, \phi, t) \right]
+\epsilon^2 \left[\rho_2(r,\zeta, \phi) +\rho'_h(r, \zeta, \phi, t)+\rho'_p(r, \zeta,\phi,t)\right].
\label{series}
\eeq

It was shown before that the first order steady term, that is $\rho_1$, can actually be set to zero.
We can approximately ignore the steady second order term $\rho_2$, as well, because even if it is not zero,
its magnitude is of the order $\epsilon^2$ as compared to the zeroth order steady term. Substituting also
the relevant terms in (75) from the explicit formulae for $\rho_1'$ and $\rho'_p$
given by formulae (\ref{all_solutions}), we get a complicated expression for the density,
including the steady base state and the $m$th Fourier mode perturbation. Rather than presenting these expressions, we recall
(see the discussion following equations (\ref{all_solutions})), that after some time only the transiently
growing terms (those proportional to T) will be dominant. These terms will be large for a rather long
time, of the order of $1/\alpha$. Thus, in this example we shall only examine the spatio-temporal behavior
of these terms. Thus we shall consider
\beq
\rho(r,\zeta,\phi,t) = \rho_0 (r,\zeta) + \epsilon^2 \left\{T e^{im\phi} \left[ \hat R^+_{(m)}(r,\zeta) e^{p^+_{(m)}T}
+\hat R^-_{(m)}(r,\zeta) e^{p^-_{(m)}T}\right]+ \rm{c.c.}\right\},
\label{fullrho}
\eeq
that is,
\beq
\rho(r,\zeta,\phi,t) = \rho_0 (r,\zeta) + \epsilon^2 \left\{T \exp{\left(-\frac{8}{5}\alpha T\right)}
 \left[ \hat R^+_{(m)}(r,\zeta)
e^{-i[(m-1) T-m\phi]}
+\hat R^-_{(m)}(r,\zeta) e^{-i[(m+1)T-m\phi]}\right]+ \rm{c.c.}\right\},
\label{finalrho}
\eeq
To obtain the surface density one has to integrate over the disk thickness, thus
\beq
\Sigma(r,\phi,t)=\Sigma_0(r) + \epsilon^2  \left\{ T \exp{\left(-\frac{8}{5}\alpha T\right)}
\left[e^{-i \theta_m^-}{\cal R}^+_m(r)+ e^{-i\theta_m^+}{\cal R}^-_m(r) + {\rm c.c.}\right]
\right\},
\label{Rtheta}
\eeq
with ${\cal R}_m^{\pm}(r)\equiv \int_{-1}^1 \hat R_{(m)}^\pm(r,\zeta) d\zeta$ and $\theta_m^\pm \equiv (m \mp 1)T-m\phi$.

Using the expression for $\rho_0$, as in equation (14) of the paper, we can easily get
\beq
\Sigma_0(r)=5^{-3/2}h^3 r^{-9/2}\int_{-1}^{1}(1-\zeta^2)^{3/2}d\zeta \approx 0.1 h^3(r) r^{-9/2},
\eeq
where $h \approx h_1 r$ (see UNRS) and $h_1$ is
a constant of order unity, very weakly dependent on the mass transfer rate and $\alpha$. This
is a particularly good approximation for $r \gg r_+$ (as we assume). In the example considered here,
we shall take $h_1=1$ for the sake of simplicity (the mass transfer rate can always be chosen accordingly).
Thus we have
\beq
\Sigma_0(r) \approx 0.1\, r^{-3/2},
\eeq
so that the surface density of the unperturbed disk increases significantly for small values of $r$.

Using equation (\ref{radial_functions}) we get
\beq
\hat R_{(m)}^{\pm }(r,\zeta) =(1-\zeta^2)^{\frac{1}{2}} [d_{0(m)}^{\pm }(r)+ d_{2(m)}^{\pm }(r)\zeta^2 + d_{4(m)}^{\pm } (r)\zeta^4],
\eeq
where, as explained before, the radial function are very complicated, albeit analytically known expressions.
Integrating over $\zeta$ we get
\beq
{\cal R}_m^\pm(r,\alpha) \equiv \int_{-1}^{1} \hat R_{(m)}^\pm(r,\zeta)d\zeta= \pi\left[ d_{0(m)}^\pm(r)+
\frac{1}{4}d_{2(m)}^\pm(r)+
\frac{1}{8}d_{4(m)}^\pm(r) \right],
\eeq
where we have explicitly reminded the $\alpha$ dependence of this quantity.

Thus the total surface density (base flow + perturbation) after a sufficiently long time, when the
algebraic term dominates (i.e., for  $t \ga  100$ time units) and before the overall exponential decay
takes over (i.e., for $t \la 1/\alpha$) can be well approximated by
\beq
\Sigma(r,\phi,t) = 0.1\, r^{-3/2} +2 \epsilon^2\left\{ T \exp{\left(-\frac{8}{5}\alpha T\right)}
\left[\cos \theta_m^-\Re({\cal R}^+_m)+\sin\theta_m^-\Im({\cal R}^+_m)+
\cos \theta_m^+\Re({\cal R}^-_m)+\sin\theta_m^+\Im({\cal R}^-_m)\right] \right\}.
\eeq
With the definitions of ${\cal R}_m^\pm$ and $\theta_m^\pm$ as above, we can now calculate the spatio-temporal
evolution of the surface density in the time interval where our approximations hold.

\subsection*{Results---example of a pattern evolution}
In what follows we shall present graphically three time snapshots of the {\em ratio} between the perturbation
of the surface density (i.e., the $\epsilon^2$ term of the above equation) and the unperturbed surface density,
$\Sigma_0$. It should be remarked that the radial functions include, in principle, two functions, $A(r)$
and $C(r)$, which are technically arbitrary, and can be only determined by the initial conditions. In the case of the
density perturbation only $A(r)$ is needed. We take it to be real, for simplicity, and to consist of
 a Gaussian peak, centered around some radius in the disk, $r_0$, in the region we wish to consider.
\beq
A(r) = g_0 e^{-(r-r_0^2)/\Delta^2},
\eeq
where the width $\Delta$ is chosen appropriately. This is done in an effort to mimic a perturbation localized in $r$.
The parameter $g_0$ determines how large the perturbation is at its peak value.
Because this expression is included in the $\epsilon^2$ term, the
resulting initial perturbation is very small, if $g_0$ is kept to be $\la 1$.\\
In this example, we follow a disk ring and are thus far enough away from the zero torque radius,
which our scaling ensured is much smaller than 1. As mentioned before, we consider an appropriate ring, because both the disk
unperturbed surface density $\Sigma_0(r)$ and the radial function ${\cal R}(r)$ exhibit a power-law decay---the former $\propto r^{-1.5}$ and the latter $\propto r^{-2}$. Thus the perturbation becomes relatively less important,
as compared to the steady quantity, for very large radii, while for small radii it may be too big.
For the sake of a clear demonstration of the transient growth we choose, as before,
$\alpha =10^{-3}$.

In the example, for which the surface density evolution is displayed in the figures, we chose only one Fourier mode ($m=2$)
for simplicity. The perturbation was introduced with a real value of $A(r)$, as given above, with $g_0=0.5$
(other choices of this parameter do not change the result significantly, as long as the parameters remain of order 1)
and with $r_0=5$, $\Delta=1$.

\begin{figure}
\begin{center}
\leavevmode \epsfysize=8cm
\epsfbox{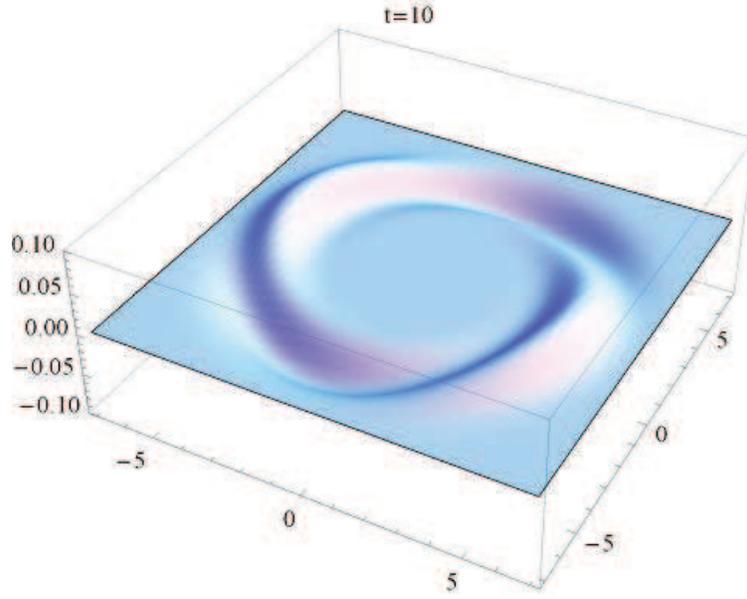}
\caption{{\small The relative perturbation in surface density, $\sigma(r,\phi,t)$, as a function of position, calculated
in a ring $1\le r \le 7$ of the disk, at a short time after the initial condition ($t=10$).
For details see the text.}}
\end{center}
\label{fig:densityten}
\end{figure}

As it can be seen in the in Figure \ref{fig:densityten}, which
displays the {\em relative} surface density perturbation, that is,
\beq
\sigma(r,\phi,t) \equiv \frac{\Sigma(r,\phi,t)-\Sigma_0(r)}{\Sigma_0(r)},
\eeq
at time $t=10$, in units of $\Omega_0^{-1}(r_\ast)$, the perturbation is still very small. Even though we cannot be sure
that at such an early time our neglecting of other terms is justified, we are confident that the
transiently growing term is still insignificant. We also verify that it has the typical $m=2$ Fourier component form.
\begin{figure}
\begin{center}
\leavevmode \epsfysize=8cm
\epsfbox{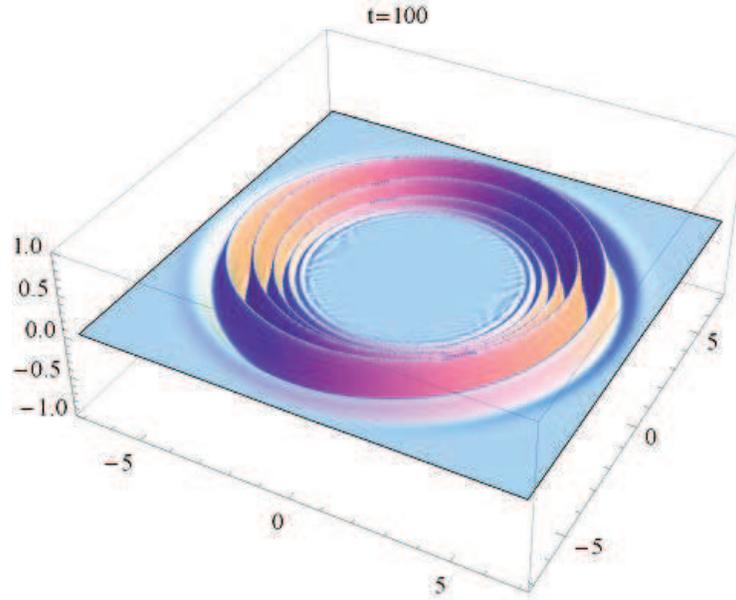}
\caption{{\small The relative perturbation in surface density, $\sigma(r,\phi,t)$, as a function of position, calculated
in a ring $1\le r \le 7$ of the disk, at time $t=100$.  The magnitude of the relative perturbation grows and its
pattern becomes more complex. For details see the text.}}
\end{center}
\label{fig:density100}
\end{figure}

\begin{figure}
\begin{center}
\leavevmode \epsfysize=8cm
\epsfbox{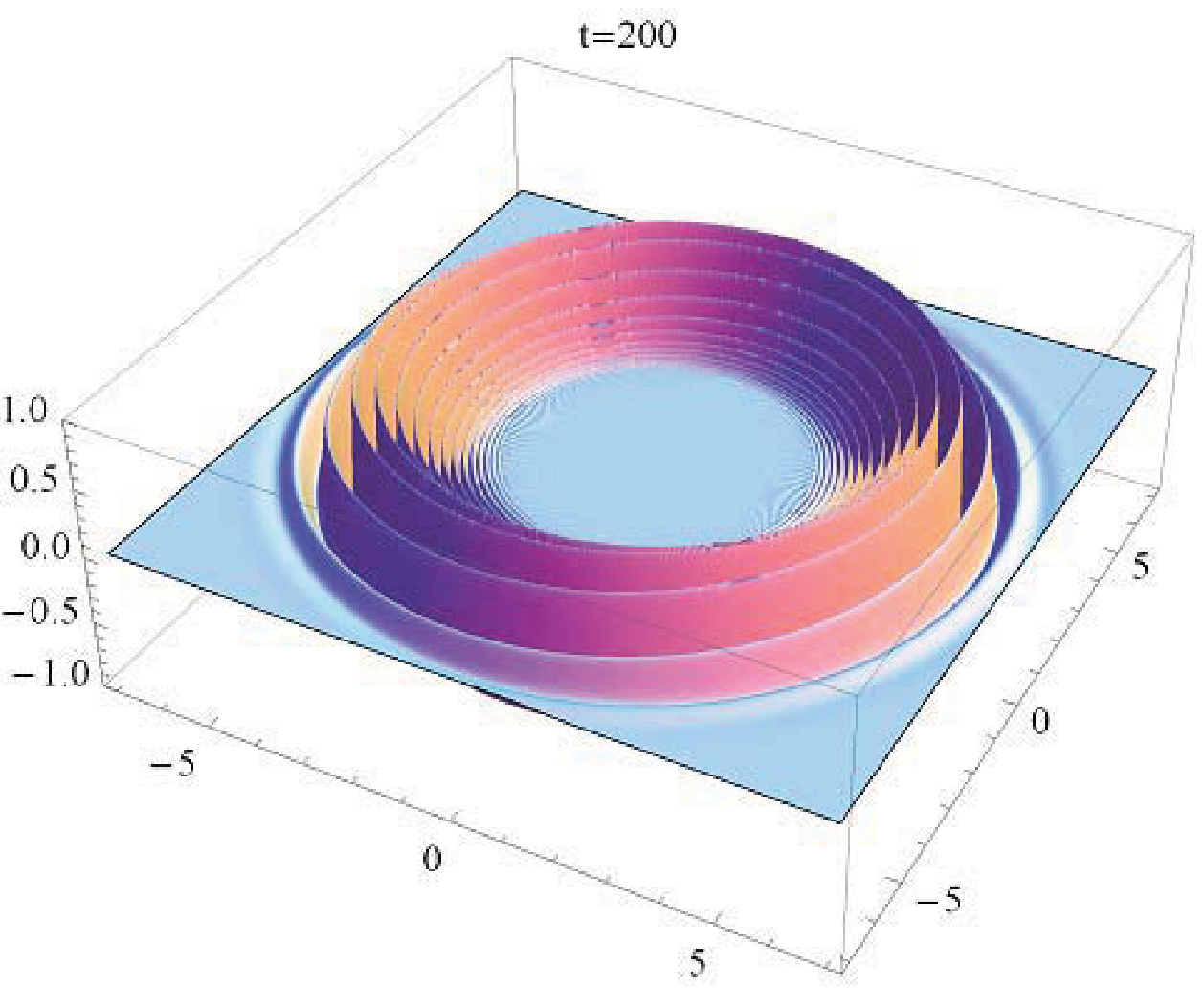}
\caption{{\small Same as Figure 6 but for time $t=200$.}}
\end{center}
\label{fig:density200}
\end{figure}

As mentioned before, we can follow the time-evolution of the surface density perturbation by calculating,
with the help of Mathematica 6, the appropriate analytical expressions.
The snapshot figures are also produced by that software.
The results for $t=100$ and $t=200$ are displayed in Figure \ref{fig:density100} and \ref{fig:density200}.
Two basic features are immediately apparent
\begin{enumerate}
\item
The absolute value of the perturbation grows with time, as it should according to the non-modal transient growth process
(see above, in the body of the paper). The growth, at these times, before the exponential decay takes over, is approximately
algebraic with $T$.
\item
Since the time variable always appears in the similarity variable $T=\Omega_0(r) t = r^{-3/2} t$, the initial perturbation
pattern (having an $m=2$ Fourier angle dependence) is being wound-up by the close to Keplerian flow, producing successive
peaks and troughs in the surface density.
\end{enumerate}

The winding of the perturbation pattern becomes rather strong for later times, giving rise to a very narrow ``radial wavelength" of the basic pattern.
The overall pattern becomes close to an
axially symmetric one by $t=100$.
Note that  $\sigma$ relates to the total time-dependent
disk surface density according to,
\beq
\Sigma(r,\phi,t)=\Sigma_0(r)\left[1+\sigma(r,\phi,t)\right].
\eeq
Thus, as long as $|\sigma|<1$, negative densities are not encountered.
Since the pattern at $t=200$ is almost axially symmetric we can display the total surface density variations
along a radial cut, that is to say, on a two dimensional plot with respect to radius for a fixed value of the azimuthal angle.
In Figure {\ref{fig:cut} we present such a cut through the disk  at $\phi=0$. The total surface density
displays a pattern resembling cylindrically symmetric ``waves", with an amplitude of the order of the
unperturbed surface density itself (also shown).
By $t=200$, the variations in the surface density
are of the same order as the unperturbed surface density, approximately $0.5\,\Sigma_0(r)\la \Sigma(r,\phi,t) \la\, 1.5 \Sigma_0(r)$.

\begin{figure}
\begin{center}
\leavevmode \epsfysize=8.cm
\epsfbox{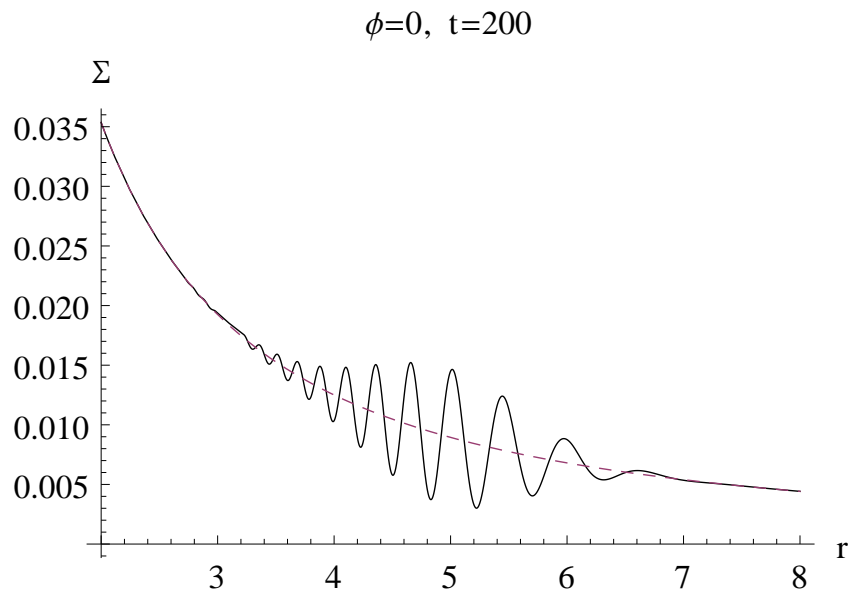}
\end{center}
\caption{{\small The perturbed surface density (in arbitrary units) as a function of $r$ (solid line). The unperturbed
surface density $\Sigma_0(r)$ is shown by the dotted line}}
\label{fig:cut}
\end{figure}

\section{Summary and discussion---A new direction toward disk turbulence?}
\label{Discussion}

In this work we considered the approximate nonlinear dynamics of a disturbed hydrodynamical viscous thin disk.
 The base flow is a Keplerian polytropic accretion disk with vertical structure (the KK analytical solution).
 By means of an asymptotic expansion in the small parameter $\epsilon$ (the ratio of the characteristic height to radius of the disk)
 we find the temporal evolution of  global non-axisymmetric perturbations. While in the first order all the variables decay, in the
 second order the perturbed density and vertical velocity display a strong transient growth.
In a short time successive peaks and troughs appear in the surface density, similar to
what was observed in the axisymmetric study of UNRS.
The fact that these structures appear for general non-axisymmetric disturbances
promotes the conjecture that this phenomenon could be commonplace in non-magnetized ADs
and this, in turn, has very interesting consequences for dynamics on the small scales. \par
To be more concrete, we start by observing that according to (\ref{Rtheta})
the surface density has the functional form $\sim Te^{-\alpha T}\sin \theta^{\pm}_m $,
where the similarity variable  $T = tr^{-3/2}$ and the radial ``wavelength"  $\theta^{\pm}_m \sim (m\mp 1) T$.  We note here two points.
Firstly, the amplitude of surface density patterns increase steadily during
the algebraic phase of the structure's response before
viscosity finally gains importance and the disturbance subsequently dies away.
The steady shortening of the acoustic pattern's  radial wavelength is a direct consequence
of the non-normal nature of the operator governing the pattern's response
and the non-separable nature of these solutions.
Secondly, the temporal decrease of the radial wavelength is enhanced for larger values of $m$, and we interpret this as the pattern getting ``wound up".  Thus, we remark that
as the pattern winds and the crenelation deepens, the amplitude of the surface density pattern
grows with time.  This latter effect is causally tied to the growing
response in the vertical velocity which comes from
shear energy being converted into vertical mechanical motions.\par

The pathways by which this
energy is fed into vertical motions can be ascertained by studying the
driving terms in the inhomogeneous operator (\ref{inho}). Referring
to the definitions given in (\ref{def_cal_F}-\ref{def_cal_J}),
we see that the effects responsible for bringing about the algebraic time dependence derive
from the operators ${\cal F}$ and ${\cal G}$ which represent, respectively,
the work done by radial compression of the radial velocity $u_1$ and
the radial gradient of $rz$ viscous stresses.  Technically speaking,
the operators ${\cal F}$ and ${\cal G}$, each having gradients with respect to
$r$, bring about factors of $T(t)$ when they act on the solution to $u_1$ (\ref{u1sol_mod}).
We note, however, that as the viscosity parameter $\alpha$
 decreases, the algebraic behavior is more strongly tied to the
compressional work rather than the work extracted from the viscous stresses, as the
latter is scaled by $\alpha$.  Nonetheless,
all perturbations that include disturbances in the angular velocity and/or
radial velocities (i.e., $u_1' , \Omega_1'$) will give rise to this algebraic behavior.
However, and although the point may be academic, we remark also that the azimuthal compressional
work of the angular velocity $\Omega_1'$ does not contribute to the algebraic growth.
This growth specifically comes from the radial compression effect $\partial_r u_1'$.
\par
The development of structure on finer scales as time goes forth also means
that terms in the asymptotic expansions formally start to break order.  This
comes about because the very same radial derivatives discussed above
no longer remain an order $\epsilon$ smaller than the $z$-derivatives
when the algebraic part of the growth becomes substantial and this is physically related
to the steady development of radial structure \cite{footnote2}.
This situation becomes
especially severe for smaller $\alpha$, as the algebraic growth persists for
timescales which are inversely proportional to the viscosity.
The breakdown of the asymptotic expansion means that the equations of motion
must be re-expanded in order to handle the evolution of
these highly wound structures with fine radial structure.  The resulting equations
will be something like the
shearing box equations \cite{GLB,UR} or some other appropriate
model set\cite{barranco,umurhan_shallow_H20,Klahr_Bodenheimer,PetersenA,PetersenB}.
\par

Fortunately, there are preliminary indications of what may happen under these circumstances.
There is recent literature devoted to exploring what
occurs on small disk scales when there is a sizable deviation from a Keplerian
flow which, in turn, is related to significant radial variations of the
surface density.  One of the effects we
expect to happen from the solution scheme employed in this work
is that the fourth order correction to the angular velocity, $\Omega_4$,
will be algebraically forced in $T$ by $\rho_2$ via the radial pressure gradient.
This can be found in the $\epsilon^4$ order expansion of (\ref{horizontal}).
As the pattern continues to wind and the crenelation deepens,
the correction $\Omega_4$ will algebraically grow with $T$, so strongly that one of several
things could possibly happen
\begin{enumerate}

\item Aside from breaking its asymptotic ordering, $\Omega_4$ would also eventually provide corrections so strong
that the \emph{composite} angular rotation profile will satisfy the Rayleigh criterion
for axisymmetric instability in many sections of the disk.
In other words, for a given small scale disk section under examination, the composite rotation
profile could conceivably develop a $1/r^2$ profile (i.e., the inviscid ``Rayleigh Line") or steeper.
This would, presumably, result in
the termination of the algebraic growth and replace it with radial transport
arising from the axisymmetric instability.
A transition may occur even before the Rayleigh line is crossed, as
Lesur and Longaretti \cite{LL} showed that a subcritical transition into
a turbulent state does exist and that the turbulent activity becomes more
vigorous as
the Rayleigh line is approached.

\item Li et al. \cite{Li_Colgate} consider the fate of linearized infinitesimal
disturbances in a vertically integrated
disk model in which there are strong radial variations of the surface density.  They
find that Rossby wave instability occurs when the surface density ``bumps"
are at least twenty percent above the mean.  Non-linear simulations show
that these non-axisymmetric instabilities turn into long-lived vortex trains
which transport significant amounts of angular momentum\cite{Li_Liska}.
Prior to the results obtained in this current work, a criticism of this proposed scenario
would have been to say that it is unrealistic
to suppose that disks are spotted with seemingly arbitrary bumps of surface density.
However, the calculation we have performed in this work shows that surface density
variations are not only common, but they can grow to large amplitude
with sufficient time.  A reference to Figure \ref{fig:cut} demonstrates how a small
perturbation can develop into sizable fluctuations of the surface density---easily
meeting the rough twenty percent minimum requirement needed to trigger
a Rossby wave instability.  Furthermore, this
transition is reported to occur as the local flow profile approaches the Rayleigh
line from the Keplerian state (see Figure 9 of Li et al. \cite{Li_Colgate}).
Preliminary calculations performed in a quasi-3D annular model of a disk\cite{umurhan_shallow_H20}
indicate that these trends are robust.  We shall detail these results
in a forthcoming work.

\end{enumerate}
We feel that the Rossby wave instability, which is a non-axisymmetric shear instability, is
likely to be strong and pronounced under these circumstances.  Of course, this assertion must be verified by further study.
Whatever the outcome may be for the smallest scales, the influence of the largest scales upon the smallest
ones ought not be ignored in the study of ADs.  The importance
of this could be deeper than previously realized and it may be
one of the reasons why ADs are such perplexing structures:
while there is an obvious separation of scales in a disk,
the solutions developed here indicate
that, if one waits long enough, the dynamics originating on
large scales invariably generates power
on the smallest scales too.  Usually this downscale cascade of
power is rationalized as happening because of nonlinear mode-mode interactions.
However, in this case it comes about due to the fundamental
inseparability of the solutions to the lowest order dynamical response.
{Nonetheless,
might it be incomplete to
examine small scale disk behavior without representing
the dynamical influence precipitating from the large scales?}\par
If secondary instabilities do develop as a result of one or more of
these processes, then
we venture to say that such radial variations of the surface density could
be either maintained or replenished due to the anomalous activity generated
by them.  This is because such activity could conceivably generate power
back onto the largest scales and, consequently, become the seed disturbances for the
large scale dynamics elucidated in this work.  This would complete a dynamical cycle
describing sustained
disk activity.  Even
if such small-to-large-scale causal connections are either absent or insignificant in disks,
one can also envisage that bursts of activity and transport can occur
in some disks simply due to random perturbations from outside
in the way discussed in Ioannou and Kakouris \cite{petros}.
Disks
certainly do not sit in isolation and periodic disturbances of them by stars
passing nearby them is likely, especially for disks found in crowded
environments like young star-forming regions.\\

\noindent The outcome of our work can be summarized as follows:
\begin{itemize}
\item Although hydrodynamical thin accretion disks are linearly stable, we find that the transient dynamics of initial 3D non-axisymmetric perturbations can
 give rise to substantial growth.  This confirms that this transient growth is not
 restricted to axisymmetric disturbances like investigated in UNRS.
\item In particular, every perturbation that disturbs the radial velocity, leads to an evolutionary
 phase in which there is algebraic growth of
 the density and vertical velocity.  After a longer time this temporal response
 gives way to an exponentially
  decaying phase wherein viscosity dominates and the perturbed quantities eventually go to zero.
\item Due to the non-normal nature of the linear operators involved, the evolution of the perturbation patterns are controlled by a similarity variable $T$, a non-separable combination
 of the radial coordinate and time.  This leads to a winding of the perturbations, producing successive peaks and troughs.  This directly contributes to the resulting algebraic growth of the other quantities discussed above as compressional work converts energy contained in the wound pattern into vertical mechanical motions.
\item For a given particular Fourier mode,  the acoustic energy associated with such a perturbation grows more for higher azimuthal number $m$ and  for smaller viscosity $\alpha$.
\item  We conjecture that as the perturbed surface density continues to wind up and deepen, secondary instabilities could arise (e.g., Rayleigh  or Rossby wave instabilities).  The development of such instabilities and their interplay with the large scale perturbed disk structures could lead to sustained turbulent activity.  These processes could contribute to the enhanced transport of angular momentum needed to match observations of ADs.
\end{itemize}
The aforementioned results should be ubiquitous for general non-magnetized accretion disks.  Therefore they are relevant for the
 understanding of accretion disks around compact objects, whether they be active galactic nuclei or close binary systems, as well as for protoplanetary disks and circumstellar disks around Be stars.
In the future we plan to investigate the possible development of secondary instabilities and the observational consequences of our findings
 in the different contexts.\\

 \noindent \textbf{Acknowledgements} P.R. is supported by the Pappalardo Postdoctoral Fellowship in Physics at MIT.
 Partial support by KBN grant N N203 381436 is acknowledged.
We are grateful to Marek Abramowicz for organizing the JPL birthday conference where this work was initiated.
We wish to thank Michael Mond for critical technical suggestions. We also like to thank Ed Bertschinger, Saul Rappaport and Bruno Coppi for valuable discussions.

%
%
%
%
\appendix
\section{The vanishing of $\Omega_1$ and $u_0$}
\label{apdx0}

Rewriting the first order equations (\ref{asol1}) and (\ref{asol2}) with the definitions
$U\equiv u_0$ and $V\equiv 2r\Omega_1$ gives
\begin{eqnarray}
\left[\left(\frac {2\alpha}{3}\right) r^3 \rho_0^{2/3}\right]\ddrv{U}{z}-
\left(\frac {2\alpha}{3}\right) z \drv{U}{z} + V=0\label{asol1ap}\\
\left[\left(\frac {2\alpha}{3}\right) r^3 \rho_0^{2/3}\right]  \ddrv{V}{z}-
\left(\frac {2\alpha}{3}\right) z \drv{V}{z}- U=0.
\label{asol2ap}
\end{eqnarray}
Substituting now the zeroth order solution $\rho_0^{2/3}=(h^2-z^2)/(5r^3)$ and rearranging
leads to
\begin{eqnarray}
U_{zz}-\frac{5z}{h^2-z^2}~ U_z +\frac{a}{h^2-z^2}~ V &=& 0\label{01}\\
V_{zz}-\frac{5z}{h^2-z^2}~ V_z -\frac{a}{h^2-z^2}~ U &=& 0,\label{02}
\end{eqnarray}
where the subscripts $z$ denote here the differentiation  with respect to $z$ and $a \equiv 15/(2\alpha)$ is a constant.

\noindent Using now $Q(z) \equiv (h^2-z^2)^{5/2}$ as an integrating factor for the first
two terms in both of the above equations, we see that
\begin{eqnarray}
\opdrv{z}\left(Q\drv{U}{z}\right)&=& -a(h^2-z^2)^{3/2}~ V\\
\opdrv{z}\left(Q\drv{V}{z}\right)&=& a(h^2-z^2)^{3/2}~ U.
\end{eqnarray}
\noindent Multiplying the first equation by $U$, the second by $V$, adding and integrating
over the domain $[-h,h]$, gives, after dropping the integrated parts,
\beq
\int_{-h}^{h}\left[\left(\drv{U}{z}\right)^2 + \left(\drv{V}{z}\right)^2\right]Q(z)dz=0.
\eeq
Because $Q(z)\neq 0$, except at $z=\pm h$ and the functions $U, V$ are bound, they must
be equal to {\em constants}. Thus, it follows from equations (\ref{01}-\ref{02}), that
$U=V=0$, except perhaps at $z=\pm h$. However, since they are bound and constant in all the domain,
they (and hence $u_0$ and $\Omega_1$) must be zero identically.
%
%
\section{Complete $\order{1}$, $\order{\epsilon}$ and $\order{\epsilon^2}$ systems}
\label{apdxcomp}
The complete equation sets in these orders will be given here.
{\em Complete} means here that it is only assumed that the zeroth order functions and
the non-perturbed portions of higher-order functions are time independent and nothing
is assumed about the axisymmetry of the solutions.

We shall also use here the notation for the full function at $i$-th order,
${\tilde f_i} = f_i(r,z,\phi)+{f_i'(r,z,\phi,t)}$.

\subsection{Zeroth order system}
\label{compO0}

The complete equations at this order are relatively simple because there is
no time dependence at this order.
\begin{eqnarray}
\Omega _0^2 &=& \frac{1}{r^3}\label{o01}\\
v_0 \rho _0 \frac{\partial \Omega _0}{\partial  z}+
 \rho _0 \Omega _0 \frac{\partial \Omega _0}{\partial  \phi}
&=&
\frac{\partial }{\partial  z}\left( \eta  \frac{\partial \Omega _0}{\partial  z}\right)
\label{o02} \\
v_0 \frac{\partial v_0}{\partial  z}+
 \Omega_0  \frac{\partial v_0}{\partial  \phi}
&=&-\frac{z}{r^3}-
\frac{\partial W_0} {\partial  z}+\frac{4}{3}\frac{1}{\rho _0}
\frac{\partial}{\partial z}\left( \eta \frac{ \partial v_0}{\partial  z}   \right)+
\frac{1}{\rho _0} \opdrv{\phi}\left( \eta  \drv{\Omega _0}{z}\right)-
\frac{2}{3}\frac{1}{\rho _0}\opdrv{z}\left(\eta \drv{\Omega _0}{\phi} \right)
\label{o03}\\
 \frac{\partial (\rho _0 v_0)}{\partial  z}+
\frac{\partial (\rho _0 \Omega _0)}{\partial  \phi} &=&0
 \label{o04}
\end{eqnarray}

\subsection{First order system}
\label{compO1}

\beq
v_0 \drv{u_0}{z}+\Omega_0\drv{u_0}{\phi}=2r \Omega_0
\tilde\Omega_1+\frac{1}{\rho_0}\opdrv{z}\left(\eta
\drv{u_0}{z}\right)
\eeq

\beq
\rho_0 \drv{\Omega'_1}{t} +
\left(\rho_0 \tilde v_1
+ v_0 {\tilde \rho_1} \right) \drv{\Omega_0}{z}+
\left(\rho_0 {\tilde \Omega_1}
+ \Omega_0 {\tilde \rho_1} \right) \drv{\Omega_0}{\phi}
+\rho_0 v_0 \drv{\tilde \Omega_1}{z}+\rho_0 \Omega_0 \drv{\tilde \Omega_1}{\phi}
=\opdrv{z}\left(\eta\drv{\tilde\Omega_1}{z}\right)-\frac{u_0 \rho_0}{r^2}\opdrv{r}\left(r^2 \Omega_0\right)
\eeq

\beqa
\drv{v'_1}{t}+u_0\drv{v_0}{r} +{\tilde v_1}\drv{v_0}{z} + v_0\drv{\tilde v_1}{z}+
{\tilde \Omega_1}\drv{v_0}{\phi}&+& v_0\drv{\tilde \Omega_1}{\phi}
+ \Omega_0\drv{\tilde v_1}{\phi} = -\drv{\tilde W_1}{z}
-\frac{2}{3 r \rho_0}\opdrv{z}\left[\eta \opdrv{r}\left(r u_0\right) \right]+\nonumber\\
+\frac{1}{r \rho_0}\opdrv{r}
\left( \eta r \drv{u_0}{z} \right)
&+&\frac{4}{3 \rho_0} \opdrv{z}\left( \eta \drv{\tilde v_1}{z}\right)
-\frac{4\tilde\rho_1}{3 \rho_0^2} \opdrv{z}\left( \eta \drv{v_0}{z}\right)
+\frac{2\tilde\rho_1} {3 \rho_0^2}\opdrv{z}\left( \eta \drv{\Omega_0}{\phi}\right)-\nonumber\\
-\frac{2}{3\rho_0}\opdrv{z}\left(\drv{\tilde\Omega_1}{\phi}\right)
&+&\frac{1}{\rho_0}\opdrv{\phi}\left(\eta\drv{\tilde\Omega_1}{z}\right)
-\frac{\tilde\rho_1}{\rho_0^2}\opdrv{\phi}\left(\eta\drv{\Omega_0}{z}\right)
\eeqa

\beq
\drv{\rho'_1}{t} +\frac{1}{r}\opdrv{r}\left(r\rho_0 u_0\right)+ \opdrv{z}\left({\rho_0 \tilde v_1}\right)
+ \opdrv{z}\left({v_0 \tilde\rho_1}\right)+ \opdrv{\phi}\left({\rho_0 \tilde \Omega_1}\right)
+ \opdrv{\phi}\left({\Omega_0 \tilde\rho_1}\right)=0
\eeq

\subsection{Second order system}
\label{compO2}
\beqa
\drv{{u'_1}}{t} + \Omega_0 \drv{{\tilde u_1}}{\phi} - 2r \Omega_0 \tilde \Omega_2
&=&-\drv{W_0}{r}+\frac{3z^2}{2 r^4}+\frac{1}{\rho_0}\opdrv{z}\left(\eta \drv{{\tilde u_1}}{z}\right)\\
\drv{\Omega'_2}{t}+ \Omega_0 \drv{\tilde \Omega_2}{\phi} +\frac{\tilde u_1}{r^2}
\opdrv{r}\left(r^2 \Omega_0\right) &=& -\frac{1}{r^3\rho_0}\opdrv{r}\left(r^3\eta\drv{\Omega_0}{r}\right)
+\frac{1}{\rho_0}\opdrv{z}\left(\eta \drv{\tilde \Omega_2}{z}\right)\\
\drv{v'_2}{t} + \Omega_0\drv{\tilde v_2}{\phi} =
 -\drv{\tilde W_2}{z}+\frac{3 z^3}{2 r^5}&+&
\frac{4}{3 \rho_0}\opdrv{z}\left(\eta \drv{\tilde v_2}{z} \right)
-\frac{2}{3\rho_0}\opdrv{z}\left(\eta \drv{\tilde \Omega_2}{\phi} \right)+
\frac{1}{\rho_0}\opdrv{\phi}\left(\eta\drv{\tilde \Omega_2}{z} \right)-\nonumber\\
&-&\frac{2}{3 r\rho_0}\opdrv{z}\left[ \eta \drv{(r \tilde u_1)}{r} \right]
+\frac{1}{r\rho_0}\opdrv{r}\left( r \eta \drv{ \tilde u_1}{z} \right)
\eeqa

\beq
\drv{{\rho'_2}}{t}+\frac{1}{r}\opdrv{r}\left(r \rho_0 {\tilde u_1} \right)+
\drv{(\rho_0 {\tilde v_2})}{z}+\opdrv{\phi}\left(\rho_0 {\tilde \Omega_2}+
\Omega_0 {\tilde \rho_2}\right)=0
\eeq
%
%
\section{First order solution, details}
\label{apdxdecay}
\subsection*{Solution}

Substituting
\beq
v'_1(r,z,\phi ,t)= \sum_{-\infty}^{\infty} \hat v_{1(m)}(r,\zeta) e^{s\Omega_0 t + \mathit{i}  m \phi } + \rm{c.c.},
\eeq
i.e., equation (\ref{Fourier}) in the linear differential equation (\ref{optau}), gives the ordinary differential equation
\begin{equation}
-\left(1 + \frac{8}{15}\alpha s_*\right)
\left[
(1-\zeta^2)\frac{\partial^2}{\partial \zeta^2} -5\zeta\frac{\partial}{\partial \zeta}
-3\frac{1+s_*^2}{1 + \frac{8}{15}\alpha s_*} \right] \hat{v}_{1(m)}(r,\zeta) = 0,
\end{equation}
where $s_*=s + \mathit{i}  m$. The previous equation  (which is similar to B.3 in UNRS, who derived it for $m=0$) is valid for any $m$ and $n=3/2$.
We can find the solution with Wolfram's Mathematica 6 (or see \cite{Math,Abra}).
Using the initial condition ${\hat v}_{1(m)}(r,0)=0$, one gets
\begin{equation}
\hat {v}_{1(m)}(r,\zeta) = \frac{A(r)}{\left(1- \zeta^2  \right)^{3/4}}\cdot
\left[
P_{_\nu}^{3/2}\left( \zeta  \right)
Q_{_\nu}^{3/2}\left(0\right)
-
Q_{_\nu}^{3/2}\left( \zeta  \right)
P_{_\nu}^{3/2}\left(0\right)
\right],
\label{solv1}
\end{equation}
where $P_\nu^\mu(\zeta)$ and $Q_\nu^\mu$ are the associated Legendre functions
of the first and second kind, respectively. $A(r)$ is the constant (in $\zeta$) of integration
and is free, depending on the initial conditions. The parameter $\nu$ relates to the
parameters of the system via,
\beq
\nu(\alpha, s_*) = \left(-\sqrt{22.5 +12 \alpha s_* }+\sqrt{90-270 s_*^2 +192 \alpha s_*} \right)/
\left(2 \sqrt{22.5+ 12 \alpha s_*}\right). \nonumber
\eeq
In addition, in order for the solutions above not to be singular at the boundaries $\zeta = \pm 1$,
$\nu$ must be equal to half integers (see next section).
\noindent The general solution to (\ref{tao14}) is the sum of the general solution to the homogeneous part and a particular solution to the inhomogeneous.
 The homogeneous solution is simply a wave---i.e., any function in the form $\zeta(\phi-\Omega_0 t)$. However, if we consider the complete
 system (\ref{tao13}-\ref{tao14}),  then we get that  $\zeta(\phi-\Omega_0 t)$ must be zero.
Therefore $\rho_1'$ is only equal to the particular solution. It can be found  by substituting
  $v_1'$  resulting from (\ref{solv1}) in (\ref{tao14}) and using the Ansatz
\beq
\rho_1'(r,z,\phi ,t) = \sum_{m=-\infty}^{\infty} \hat \rho_{1(m)}(r,\zeta) e^{ s \Omega_0 t + \mathit{i}  m \phi } + \rm{c.c.},
\eeq
Then, for a given $m$, we find

\begin{eqnarray}
 \rho_1^{(m)}(r,\zeta)& =& \frac{1}{10 \sqrt{5} r^{3/2}  s_*
 \left(-1+\zeta ^2\right)^{3/4}}  (2 \nu-1) A(r) h(r)\times \\
& &\sqrt{\frac{\left(1-\zeta ^2\right) h(r)^2}{r^3}} \left[\zeta  P_\nu^{3/2}(\zeta) Q_{\nu}^{3/2}(0)-
P_{\nu+1}^{3/2}(\zeta) Q_{\nu}^{3/2}(0)+P_{\nu}^{3/2}(0) \left(-\zeta  Q_\nu^{3/2}(\zeta)+Q_{\nu+1}^{3/2}(\zeta)\right)\right].
\nonumber
\label{solrho1}
\end{eqnarray}

The eigenvalues, $s$, for any $m$, will follow from the boundary conditions on $\zeta$ (see below).

\subsection*{Vertical boundary conditions}

We assume that the lagrangian pressure perturbation vanishes at $z=\pm h(r)$

\begin{equation}
\frac{dP}{dt}=0=(\frac{\partial}{\partial t}+\vec{v}\cdot\vec{\nabla})P=0 , \ \ \ {\rm at} \ \ z = \pm h(r) .
\label{eqBC1}
\end{equation}

Using the polytropic equation and the continuity equation, (\ref{eqBC1}) can be simplified to

\begin{equation}
P \vec{\nabla} \cdot \vec{v}=0 , \ \ \ {\rm at} \ \ z = \pm h(r)
\label{eqBC2}
\end{equation}

At this order, since $u_0=\Omega_1=0$, it reduces to

\begin{equation}
P_0(r,z)\frac{\partial v_1'}{\partial z}=0 , \ \ \ {\rm at} \ \ z = \pm h(r) .
\label{eqBCcaseB}
\end{equation}

We substitute the Ansatz for $v_1'$ and the solution \ref{solv1} into (\ref{eqBCcaseB}) and consider
the limit for $\zeta \rightarrow \pm 1$.
Therefore the boundary conditions (for $n>0$ an integer or half odd integer) resulting from regularity conditions at the disk upper and lower edges
are satisfied for $\nu= 2 k+5/2$, where $k \geq 0$ is an integer or $k=-3/2$.
Following the discussion in UNRS B.1 we can consider the fundamental mode (k=0) and so find the
 two fundamental eigenvalues ($s_{(m)}^{\pm}$), that lead to
a typical oscillatory decay. The values of $\alpha$ which we consider in this work ($\alpha \ll 1$)
do not allow  a non-oscillatory decay---see (\ref{eigen1}).
The actual solution for the fundamental mode ($k=0$) becomes

\beq
v'_1(r,z,\phi ,t)= \sum_{-\infty}^{\infty} \hat v_{1(m)}(r,\zeta)(S_+ e^{s_{(m)}^+ T + \mathit{i}  m \phi } + S_- e^{s_{(m)}^- T + \mathit{i}  m \phi })+c.c.,
\eeq

and

\beq
\rho'_1(r,z,\phi ,t)= \sum_{-\infty}^{\infty} \hat \rho_{1(m)}(r,\zeta)(S_+^\rho e^{s_{(m)}^+ T + \mathit{i}  m \phi } + S_-^\rho e^{s_{(m)}^- T + \mathit{i}  m \phi })+c.c.
\eeq

\subsection*{Surface stress conditions}

It is easy to see that the vanishing stress conditions are satisfied on the surface

\begin{equation}
\lim_{z\rightarrow h} \
\eta  {{\bf \hat n}}\cdot{\vec{\nabla}}u   \rightarrow 0 \  \  {\rm and} \  \   \lim_{z\rightarrow h} \
\eta r {{\bf \hat n}}\cdot{\vec{\nabla}}\Omega   \rightarrow 0,
\end{equation}
in which $\hat n=\hat z$ to lowest order.

%
%

\section{Second order solution---details}
\label{appdxgrowth}

As detailed in the text,(\ref{Frob}) holds for the $m$-th Fourier component
of the function $u_1'(r,z,\phi)$, see (\ref{u1sol}). We rewrite it here for convenience.
\beq
\left\{\left[ \frac{1}{5}\alpha(1-\zeta^2)\frac{\partial^2}{\partial\zeta^2}-\alpha\zeta\frac{\partial}
{\partial\zeta} -\frac{3}{2}p_*\right]^2 +\frac{9}{4}\right\} \hat u_{1(m)}(r,\zeta)=0,
\label{Frobapp}
\eeq
where $\zeta = z/h(r)$ and $p_*=p+ \mathit{i}  m$ is an eigenvalue.

A similar equation is also satisfied by the Fourier components of $\Omega_2'(r,z,\phi)$, see
(\ref{om2sol}).
To solve the above equation we assume a truncated series expansion, which for a general
$k\geq 0$ reads
\begin{equation}
\hat u_{1(m)} = \hat u^{(k)}_{(m)} = \sum_{j=0}^{k+1}A_j(r) \zeta^{2(k+1)-2j},
\label{solution_u}
\end{equation}
 We consider the fundamental mode $k=0$ and the relative eigenvalue $p$, as above,
and obtain
\begin{equation}
\hat u_1 = \hat u_1^{(0)} =
A(r) \zeta^2 + B(r),
\label{fundamental_u}
\end{equation}
where we have dropped the $(m)$ subscript for economy of notation.

Inserting  (\ref{fundamental_u}) into (\ref{Frobapp}) leads to
\beq
\zeta^2  \left(225 \left(1+p_*^2\right) + 144 \alpha  (5 p_*+4 \alpha ) \right) A(r) +  \left(225 \left(1+p_*^2\right) B(r) - 24 \alpha  (5 p_*+4 \alpha ) A(r) \right)=0.
\eeq

\noindent We set to zero the coefficient of $\zeta ^2$ to find $p_* $, and therefore $p$, and the coefficient of $\zeta ^0$ to find $B(r)$

\begin{equation}
p_{(m)}^\pm = - \frac{8}{5}\alpha -i m \pm i ,
\qquad
\frac{B(r)}{A(r)} = -\frac{1}{6}.
\label{sigma_u}
\end{equation}

These modes are decaying oscillations.
We  then substitute $v_p'$, $z=\zeta h$,$\hat u_1$ as in (\ref{fundamental_u}) and the fundamental mode
 $\hat \Omega_2 = \hat \Omega_2^{(0)} =Q(r)(z^2/h^2 - 1/6)$ in (\ref{inho}) we get ($n=3/2$)

\begin{eqnarray}
& &\hat v_{(m)}^{(0)}+(p+ \mathit{i}  m)^2 \hat v_{(m)}^{(0)}+2 (p+ \mathit{i}  m) \hat V_{(m)}^{(0)}+\frac{5}{3} \zeta \drv{ \hat v_{(m)}^{(0)}}{\zeta}-\frac{1}{3} \ddrv{\hat v_{(m)}^{(0)}}{\zeta}+\frac{1}{3} \zeta ^2 \ddrv{ \hat v_{(m)}^{(0)}}{\zeta}+\nonumber \\
& &t r^{-3/2}\left(\hat V_{(m)}^{(0)}+(p+ \mathit{i}  m)^2 \hat V_{(m)}^{(0)}+\frac{5}{3} \zeta  \drv{\hat V_{(m)}^{(0)}}{\zeta}-\frac{1}{3} \ddrv{\hat V_{(m)}^{(0)}}{\zeta}+\frac{1}{3} \zeta ^2 \ddrv{\hat V_{(m)}^{(0)}}{\zeta}\right) = \nonumber \\
&=&t \zeta  \left(-\frac{7 (p+ \mathit{i}  m) A h}{6 r^{5/2}}-\frac{(p+ \mathit{i}  m)^2 \alpha  A h}{45 r^{5/2}}+m \left(\frac{7 i A h}{6 r^{5/2}}+\frac{i (p+ \mathit{i}  m) \alpha  A h}{45 r^{5/2}}\right)\right)+\nonumber  \\
&+& t \zeta ^3 \left(\frac{2 (p+ \mathit{i}  m) A h}{r^{5/2}}-\frac{8 (p+ \mathit{i}  m)^2 \alpha  A h}{15 r^{5/2}}+m \left(-\frac{2 i A h}{r^{5/2}}+\frac{8 i (p+ \mathit{i}  m) \alpha  A h}{15 r^{5/2}}\right)\right) \nonumber \\
&+& \zeta ^3 \left(\frac{14 A h}{3 r}+\frac{64 (p+ \mathit{i}  m) \alpha  A h}{45 r}+m \left(\frac{8 i \alpha  A h}{15 r}-\frac{4}{3} i h Q+\frac{16}{45} i (p+ \mathit{i}  m) \alpha  h Q\right)-\frac{4}{3} h A'+\frac{16}{45} (p+ \mathit{i}  m) \alpha  h A'\right)+ \nonumber\\
&+&\zeta  \left(-\frac{49 A h}{18 r}-\frac{217 (p+ \mathit{i}  m) \alpha  A h}{135 r}+m \left(\frac{i \alpha  A h}{45 r}+\frac{7}{9} i h Q+\frac{2}{135} i (p+ \mathit{i}  m) \alpha  h Q\right)+\frac{7}{9} h A'+\frac{2}{135} (p+ \mathit{i}  m) \alpha  h A' \right) +\nonumber\\
&+& \zeta \left(2 A h'+\frac{4}{3} (p+ \mathit{i}  m) \alpha  A h'\right)
\label{eq_app}
\end{eqnarray}

\subsection*{Vertical boundary Conditions}

At the second order (\ref{eqBC2}) becomes

\beq
\frac{P_0}{r}(u_1'+r \frac{\partial u_1'}{\partial r}) + P_0 ( \frac{\partial v_2'}{\partial z}+ \frac{\partial \Omega_2'}{\partial \phi})+P_1'  \frac{\partial v_1'}{\partial z}=0 , \ \ \ {\rm at} \ \ z = \pm h(r) .
\label{eqBCcaseC}
\eeq

The first two terms satisfy this condition since $P_0=\left(\frac{h^2-z^2}{5 r^3} \right)^{5/2}$ is zero
 at $z = \pm h$ and $u_1'$, $v_2'$ and $\Omega_2'$ and their derivatives are finite.
The last term vanishes as well, for $k=-3/2$ or integer $\geq 0$.
Therefore the BC is satisfied also at this order.

\section{Mass accretion rate}
We want

\begin{equation}
\ r\int_{0}^{2 \pi} d\phi \int_{-h}^{h} {\rho  u} dz = -\dot{M}=const.
\end{equation}

If we expand $\rho$ and $u$, we get different equations at different orders.
The time-independent parts are

\begin{equation}
\int_{0}^{2 \pi} d\phi \int_{-h}^{h} {r \rho_0  u_1} dz =  -\dot{M_1}=const. \qquad \int_{0}^{2 \pi} d\phi \int_{-h}^{h} {r \rho_2  u_1} dz =  -\dot{M_2}=const.
\end{equation}

Therefore $\dot{M}= \epsilon \dot{M_1}+\epsilon^3 \dot{M_2}$.

For the time-dependent part we get

\begin{equation}
\int_{0}^{2 \pi} d\phi \int_{-h}^{h} {r \rho_0  u_1'} dz =  0, \qquad \int_{0}^{2 \pi} d\phi \int_{-h}^{h} {r \rho_1'  u_1} dz =  0.
\end{equation}

\begin{equation}
\int_{0}^{2 \pi} d\phi \int_{-h}^{h} {r \rho_2'  u_1} dz =  0, \qquad \int_{0}^{2 \pi} d\phi \int_{-h}^{h} {r \rho_2  u_1'} dz =  0.
\end{equation}

However

\begin{equation}
\int_{0}^{2 \pi} d\phi \int_{-h}^{h} {r \rho_1'  u_1'} dz \neq  0, \qquad \int_{0}^{2 \pi} d\phi \int_{-h}^{h} {r \rho_2'  u_1'} dz \neq  0.
\end{equation}

Indeed in both cases the function that has to be integrated is even in $z$ and is not $2 \pi-$ periodic in $\phi$.
We can satisfy the first equation by setting  $ \rho_1'\equiv 0$, but the second equation does not vanish and
gives a correction to $\dot M$ that grows and subsequently decays like $\sim \epsilon^3 T e^{-\frac{16}{5} \alpha T }$. This is
 a fluctuation of the order of $\epsilon^3$ on a quantity ($\dot{M}$) of the order of $\epsilon$. It can be neglected for
 $\epsilon$ sufficiently small, that is for $\alpha$ not too small.

\section{Energy-details}

The particular solutions $v_p'$ and $\rho_p'$ are both in the form $f_p'=A_{m} e^{\Gamma_{m}}+B_{m} e^{G_{m}}+{\rm c.c.}$, see (\ref{sol_v}).
Let now $A_{m}= A_{\rm R} + \mathit{i} A_{\rm I}$ and, likewise,  $\Gamma_{m}= \Gamma_{\rm R} + \mathit{i}
\Gamma_{\rm I}$,  $B_{m}= B_{\rm R} + \mathit{i} B_{\rm I}$, $G_{m}= G_{\rm R} + \mathit{i} G_{\rm I}$, where
$A_{\rm R},A_{\rm I},\Gamma_{\rm R}$ , $\Gamma_{\rm I}$, $B_{\rm R}, B_{\rm I}, G_{\rm R}$  and  $G_{\rm I}$ are all real quantities.
Then we have

\begin{equation}
{\rm Re}(f_p')^2= e^{2 \Gamma_{\rm R}}({\cal F}_a +{\cal F}_b + {\cal F}_c),
\end{equation}
where
\begin{equation}
{\cal F}_a\equiv
2 A_{\rm I}^2 + 2 A_{\rm R}^2 + 2 B_{\rm I}^2 + 2 B_{\rm R}^2 +
(4 A_{\rm I} B_{\rm I} + 4 A_{\rm R} B_{\rm R}) \cos{(2 T)} + (- 4 A_{\rm I} B_{\rm I} + 4 A_{\rm R} B_{\rm R})
\cos{( \Gamma_{\rm I}+ G_{\rm I})},\nonumber\\
\end{equation}
\begin{equation}
{\cal F}_b\equiv
 - 2 (B_{\rm I}^2+B_{\rm R}^2)  \cos{(2 G_{\rm I})} +  (-2 A_{\rm I}^2 + 2 A_{\rm R}^2) \cos{(2 \Gamma_{\rm I})}
 + 4 (A_{\rm R} B_{\rm I}-4 A_{\rm I} B_{\rm R} ) \sin{(2 T)}, \nonumber\\
\end{equation}
\begin{equation}
 {\cal F}_c\equiv- 4 (A_{\rm R} B_{\rm I} +A_{\rm I} B_{\rm R}) \sin{( \Gamma_{\rm I}+ G_{\rm I})}-
 4 B_{\rm I} B_{\rm R} \sin{(2 G_{\rm I})} - 4 A_{\rm I} A_{\rm R} \sin{(2\Gamma_{\rm I})}.\nonumber
\label{realsq}
\end{equation}
$v_p'$ and $\rho_p'$ have the same phases, i.e., $\Gamma_{\rm R}=G_{\rm R}= - (8/5) \alpha T$,
$\Gamma_{\rm I}= T (1-m)+ m \phi$ and $\Gamma_{\rm I}= T (1+m)+ m \phi$.
The $\phi$ integral of the quantity in (\ref{realsq}) can be simplified using
trigonometric relations

\begin{equation}
\int_{0}^{2 \pi}\frac{1}{2} {\rm Re}(f_p')^2 d\phi= 2 \pi e^{-\frac{16}{5}\alpha T }
\left[ A_{\rm R}^2  +A_{\rm I}^2 +B_{\rm R}^2  +B_{\rm I}^2 + 2 (A_{\rm I} B_{\rm I}
 +A_{\rm R} B_{\rm R}) \cos{(2 T)} +
+ 2 (A_{\rm R} B_{\rm I} +A_{\rm I} B_{\rm R}) \sin{(2 T)}\right],
\label{eq:enintm}
\end{equation}

For $m =0$  the above equation reduces to

\begin{equation}
\int_{0}^{2 \pi}\frac{1}{2} {\rm Re}(f_p')^2 d\phi=4 e^{-\frac{16 T \alpha }{5}} \pi
\left[(A_{\rm R}+B_{\rm R}) \cos(T)-(A_{\rm I}-B_{\rm I}) \sin(T)\right]^2
\label{eq:enint0}
\end{equation}

We now write equation (\ref{eq:en}) in the form (F1) and proceed with the vertical integration,
remembering that in the fundamental mode ($k=0$)
we had for the radial and angular perturbations
$\hat u_{1(m)}(r,\zeta) = A(r)\left(\zeta^2 - \frac{1}{6} \right)$ and
 $\hat \Omega_{2(m)}(r,\zeta) =C(r)\left(\zeta^2 - \frac{1}{6}\right)$,
with the radial functions $A(r)$ and $C(r)$ free.

Moreover, we notice that $ h(r) \rightarrow (2 \Lambda)^{1/6} r=c_1 r$ for $r>>r_*$, with
$c_1$  depending on $\alpha$ and on the mass flux. However it multiplies every coefficient and
therefore we can set it to one without loss of generality.
After the integration in the vertical direction, we obtain $ {\cal E}_r$ in the form

\begin{equation}
{\cal E}_r(r,T;\alpha,m) = e^{-\frac{16}{5}  \alpha T} F(r;\alpha,m,\cos{2 T},\sin{2 T}).
\end{equation}

$F$ is a known analytical function: we shall not write it out explicitly for space considerations.


\begin{thebibliography}{}
\bibitem{footnote1}{}
        {The quality of the KK solution
has been shown to be virtually unaffected if one relaxes the polytropic assumption
and use, instead, a more realistic model for the disk's thermal structure \cite{RG}. }
\bibitem{footnote2}{}
{In other words,
since the operator $\partial_r$ includes $Tr^{-5/2}\partial_T$ as one of its terms,
the term acted on by it produces a term which will eventually break order.}
\end{thebibliography}
\end{document}